\documentclass[12pt,twoside]{article} 
\usepackage{pdproc} 
\usepackage{graphicx}
\usepackage{latexsym}
\usepackage{hyperref}

  \newcommand{\snu}{\ensuremath{\tilde{\nu}}}

  \def\ket#1{| #1\rangle}

  \newcommand{\gev}{\hbox{ GeV}}
  
  \newcommand{\ev}{\hbox{ eV}}

  \newcommand{\tev}{\hbox{ TeV}}

  \newcommand{\s}{\hbox{ s}}
  
  \newcommand{\cm}{\hbox{ cm}}

  \newcommand{\mpc}{\hbox{ Mpc}}

  \newcommand{\km}{\hbox{ km}}

  \newcommand{\y}{\hbox{ y}}

  \def\ltap{\mathop{\raisebox{-.4ex}{\rlap{$\sim$}} 
  \raisebox{.4ex}{$<$}}}
  \def\gtap{\mathop{\raisebox{-.4ex}{\rlap{$\sim$}} 
  \raisebox{.4ex}{$>$}}}

  \newcommand{\cfrac}[2]{\textstyle \frac{#1}{#2}}

  \newcommand{\m}{\hbox{ m}}

  \def\slashii#1{\setbox0=\hbox{$#1$}             
     \dimen0=\wd0                                 
     \setbox1=\hbox{\sl/} \dimen1=\wd1            
     \ifdim\dimen0>\dimen1                        
	\rlap{\hbox to \dimen0{\hfil\sl/\hfil}}   
	#1                                        
     \else                                        
	\rlap{\hbox to \dimen1{\hfil$#1$\hfil}}   
	\hbox{\sl/}                               
     \fi}                                         %
  \def\slashiii#1{\setbox0=\hbox{$#1$}#1\hskip-\wd0\hbox to\wd0{\hss\sl/\/\hss}}
  \newsavebox{\savepar}

  \def\slashii#1{\setbox0=\hbox{$#1$}             
     \dimen0=\wd0                                 
     \setbox1=\hbox{\sl/} \dimen1=\wd1            
     \ifdim\dimen0>\dimen1                        
	\rlap{\hbox to \dimen0{\hfil\sl/\hfil}}   
	#1                                        
     \else                                        
	\rlap{\hbox to \dimen1{\hfil$#1$\hfil}}   
	\hbox{\sl/}                               
     \fi}                                         %
\catcode`\@=11
\def\ps@ppt{\def\@oddhead{\qquad \textsc{Cosmic and Relic Neutrinos}\hfil \thepage\qquad}
\def\@evenhead{\qquad\thepage \hfil \textsc{Chris Quigg} \qquad}
\def\@oddfoot{}\def\@evenfoot{}}    
\catcode`\@=12
\begin{document}
\textsf{astro-ph/0603372} \hfill    \textsf{FERMILAB--CONF--06/029--T }\\ 
    \phantom{M}

\renewcommand{\thefootnote}{\alph{footnote}}
  
\title{
EXTREMELY HIGH ENERGY COSMIC NEUTRINOS\\
 AND RELIC NEUTRINOS}

\author{ CHRIS QUIGG}

\address{ Fermi National Accelerator Laboratory \\
  P.O. Box 500,
 Batavia, Illinois 60510 USA
\\ and \\
 CERN, CH-1211 Geneva 23, Switzerland  \\ {\rm E-mail: quigg@fnal.gov} }




\abstract{I review the essentials of ultrahigh-energy neutrino interactions, 
show how neutral-current detection and flavor tagging can enhance the scientific 
potential of neutrino telescopes, and sketch new studies on neutrino encounters 
with dark matter relics and  on gravitational lensing of neutrinos.}
   
\normalsize\baselineskip=15pt

\section{Neutrino Observatories: Expectations}
An early goal of the next generation of neutrino telescopes will be to
detect the flux of cosmic neutrinos that we believe will begin to show
itself above the atmospheric-neutrino background at energies of a few
TeV. A short summary of the science program of these instruments is to
prospect for cosmic-neutrino sources, to characterize the sources, to
study neutrino properties, and to be sensitive to new phenomena in
particle physics.  The expected sources include active galactic nuclei
(AGN) at typical distances of roughly $100\mpc$.\footnote{$1\mpc
\approx 3.1 \times 10^{22}\m$.} If neutrinos are produced there in the 
decay of pions created in $pp$ or $p\gamma$ collisions, then we 
anticipate---at the source---equal numbers of neutrinos and 
antineutrinos, with a flavor mix $2 \gamma + 2 \nu_{\mu} + 
2\bar{\nu}_{\mu} + 1 \nu_e + 1 \bar{\nu}_e$, provided that all pions 
and their daughter muons decay. I  denote this standard flux at 
the source by
$\Phi^{0}_{\mathrm{std}} = \{\varphi_{e}^{0} = \cfrac{1}{3},
    \varphi_{\mu}^{0} = \cfrac{2}{3}, \varphi_{\tau}^{0} =
    0\}$.

We expect that a neutrino observatory with an instrumented volume of
$1\km^{3}$ will be able to survey the cosmic-neutrino flux over a broad
range of energies, principally by detecting the charged-current
interaction $(\nu_{\mu},\bar{\nu}_{\mu})N \rightarrow
(\mu^-,\mu^+)+\hbox{anything.}$ Important open questions are whether we
can achieve efficient, calibrated $(\nu_e, \bar{\nu}_e)$ and
$(\nu_{\tau},\bar{\nu}_{\tau})$ detection, and whether we can record
and determine the energy of neutral-current events.  One of my aims in
this talk will be to illustrate how adding these capabilities will enhance
the scientific potential of neutrino observatories.

The cross section for deeply inelastic scattering on an isoscalar 
nucleon may be written in terms of the Bjorken scaling variables 
$x = Q^2/2M\nu$  and $y = \nu/E_\nu$ as
\begin{equation}
    \frac{d^2\sigma}{dxdy} = \frac{2 G_F^2 ME_\nu}{\pi} 
    {\left( 
    \frac{M_W^2}{Q^2 + M_W^2} \right)^{\!2}} \left[xq(x,Q^2) + x 
    \bar{q}(x,Q^2)(1-y)^2 \right]  \;,
    \label{eqn:sigsig}
\end{equation}
where $-Q^2$ is the invariant momentum transfer between
the incident
neutrino and outgoing muon, $\nu = E_\nu - E_\mu$ is the energy loss in
the lab (target) frame, $M$ and $M_W$ are the nucleon and
intermediate-boson masses, and $G_F = 1.16632 \times 10^{-5}\gev^{-2}$ is
the Fermi
constant. The parton densities are
\begin{eqnarray}
q(x,Q^2) & = & \frac{u_v(x,Q^2)+d_v(x,Q^2)}{2} +
\frac{u_s(x,Q^2)+d_s(x,Q^2)}{2} \nonumber\\ & & + s_s(x,Q^2) + b_s(x,Q^2)  
 \\[12pt]
	\bar{q}(x,Q^2) & = & \frac{u_s(x,Q^2)+d_s(x,Q^2)}{2} + c_s(x,Q^2) + 
	t_s(x,Q^2),\nonumber 
\end{eqnarray}
where the subscripts $v$ and $s$ label valence and sea contributions,
and $u$, $d$, $c$, $s$, $t$, $b$ denote the distributions for various
quark flavors in a {\em proton}.  The $W$-boson propagator, which has a
negligible effect at low energies, modulates the high-energy cross
section and has important consequences for the way the cross section is
composed.

I was drawn to this problem by the
observation\cite{Andreev:1979cp} that the $W$-boson propagator squeezes
the significant contributions of the parton distributions toward
smaller values of $x$ with increasing energy.  There the QCD-induced
growth of the small-$x$ parton distribution enhances the high-energy
cross section.  This stands in contrast to the familiar effect of QCD
evolution at laboratory energies, which is to diminish the total cross
section as the valence distribution is degraded at high values of
$Q^{2}$. At that moment, my colleagues and I had developed for our 
study of supercollider physics\cite{Eichten:1984eu} the first all-flavor set of parton 
distributions appropriate for applications at small $x$ and large 
$Q^{2}$, so I had in my hands everything needed for a modern 
calculation of the charged-current cross section at ultrahigh 
energies. In a sequence of works on the 
problem,\cite{Quigg:1986mb,Reno:1987zf,Gandhi:1995tf,Gandhi:1998ri} we 
have tracked the evolving experimental understanding of parton distributions 
and investigated many facets of ultrahigh-energy neutrino interactions.

\begin{figure}[tb]
    \centerline{\includegraphics[height=5.4cm]{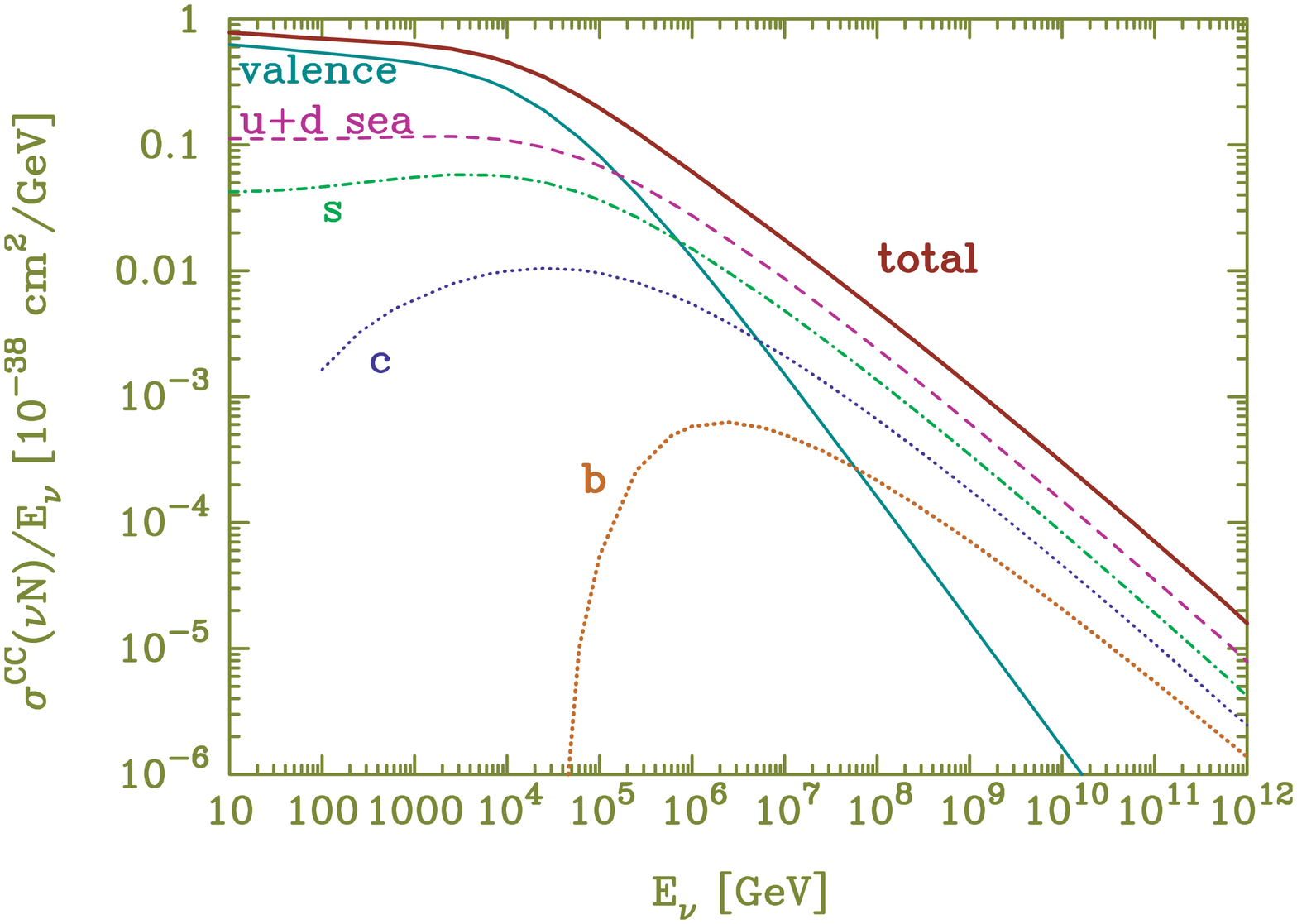}\quad
    \includegraphics[height=5.3cm]{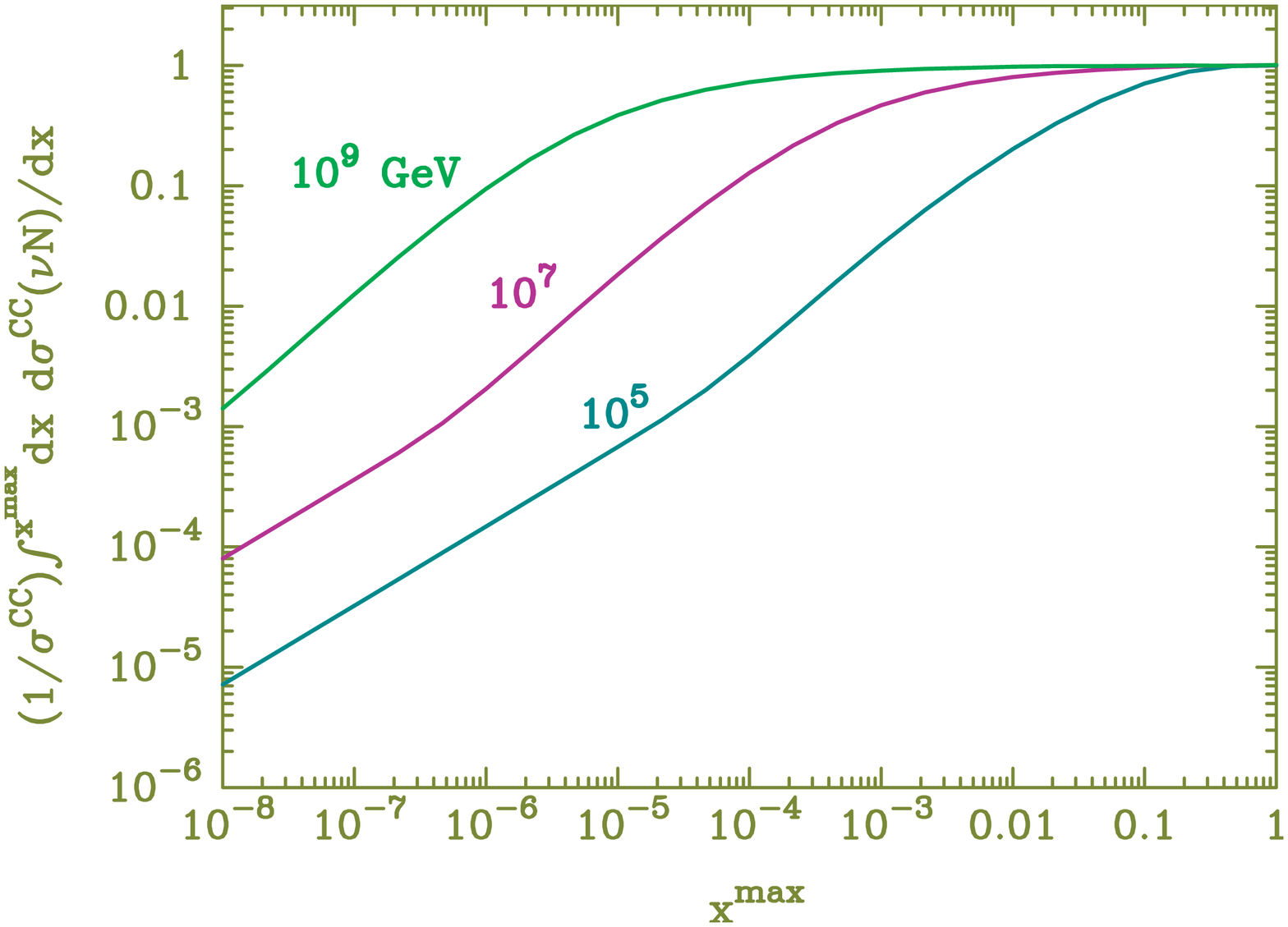}}
\caption{Left panel: Components of the $\nu N$ charged-current cross
section as functions of the neutrino energy for the CTEQ3
distributions. Right panel: Integral cross section
$(1/\sigma)\int_{0}^{x^{\mathrm{max}}}dx\:d\sigma/dx$ for the
charged-current reaction $\nu_{\mu}N \rightarrow \mu^{-}+\hbox{
anything}$ at $E_{\nu}= 10^{5}, 10^{7},\hbox{ and }10^{9}\gev$.  As the
neutrino energy increases, the dominant contributions come from smaller
values of $x$.{\protect\cite{Gandhi:1995tf}}.}
\label{fig:flavors}
\end{figure}
Let us recall some of the principal lessons.  The left panel of
Figure~\ref{fig:flavors} shows the contributions of various parton
species to the total charged-current cross section, as a function of
energy.  Note that the valence contribution, which dominates at
laboratory energies, becomes negligible above about $10^{16}\ev$,
whereas strange- and charm-quark contributions become significant.  The
cross section integrals in the right panel of Figure~\ref{fig:flavors}
illustrate the increasing prominence of small values of $x$ as
the neutrino energy increases.  At $E_{\nu} = 10^{5}\gev$, nearly all of
the cross section comes from $x \gtap 10^{-3}$, but by $E_{\nu} =
10^{9}\gev$, nearly all of the cross section lies below $x = 10^{-5}$, 
where we lack direct experimental information. Reno has given 
a comprehensive review of small-$x$ uncertainties and the possible influence of 
new phenomena on the total cross section.\cite{Reno:2004cx}

\begin{figure}[tb]
    \centerline{\includegraphics[width=8cm]{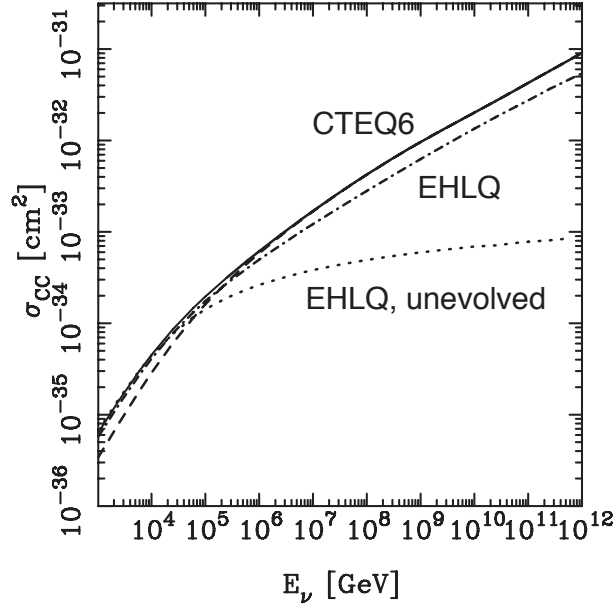}}
    \caption{The solid curve shows the charged-current $\nu N$ cross section 
    calculated using the CTEQ6 parton distributions;\protect\cite{Pumplin:2002vw}
    the dash-dotted line shows the situation in 1986, using Set~2 of 
    the EHLQ parton distributions.\protect\cite{Eichten:1984eu} The dotted curve shows the energy 
    dependence of the cross section without QCD evolution, i.e., with 
    the EHLQ distributions frozen at $Q^{2} = 5\gev^{2}$.\protect\cite{Reno:2004cx}}
\label{fig:ctgeq6}
\end{figure}

Figure~\ref{fig:ctgeq6} compares our first calculation, using the 1984
EHLQ structure functions, with the recent CTEQ6 parton
distributions.\cite{Pumplin:2002vw} At the highest energies plotted,
the cross section is about $1.8 \times$ our original estimates, because
today's parton distributions rise more steeply at small $x$ than did
those of two decades ago.  HERA measurements have provided the decisive
new information.\cite{Adloff:2003uh} At $10^{12}\gev$, the QCD
enhancement of the small-$x$ parton density has increased the cross
section sixty-fold over the parton-model prediction without evolution.
HERA measurements of the charged-current reaction $e p \to \nu +
\hbox{anything}$ at an equivalent lab energy near $40\tev$ observe the
damping due to the $W$-boson propagator and agree with standard-model
cross sections.\cite{Ahmed:1994fa}
\begin{figure}[b!]
    \centerline{\includegraphics[height=6.5cm]{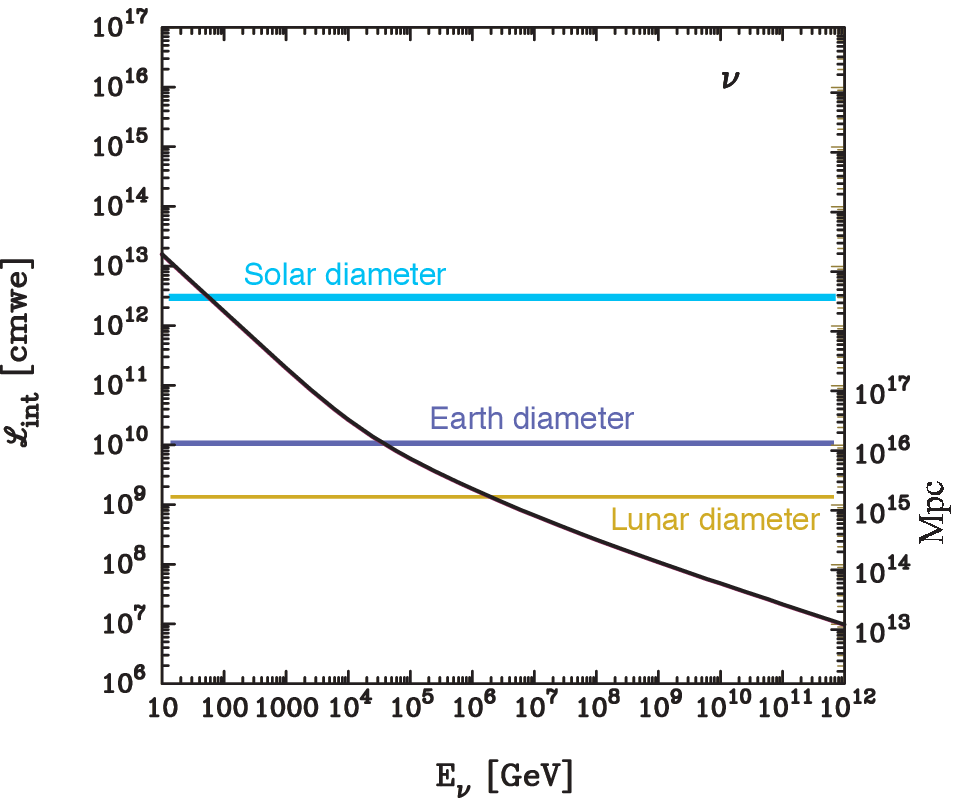}\quad
\includegraphics[height=6.5cm]{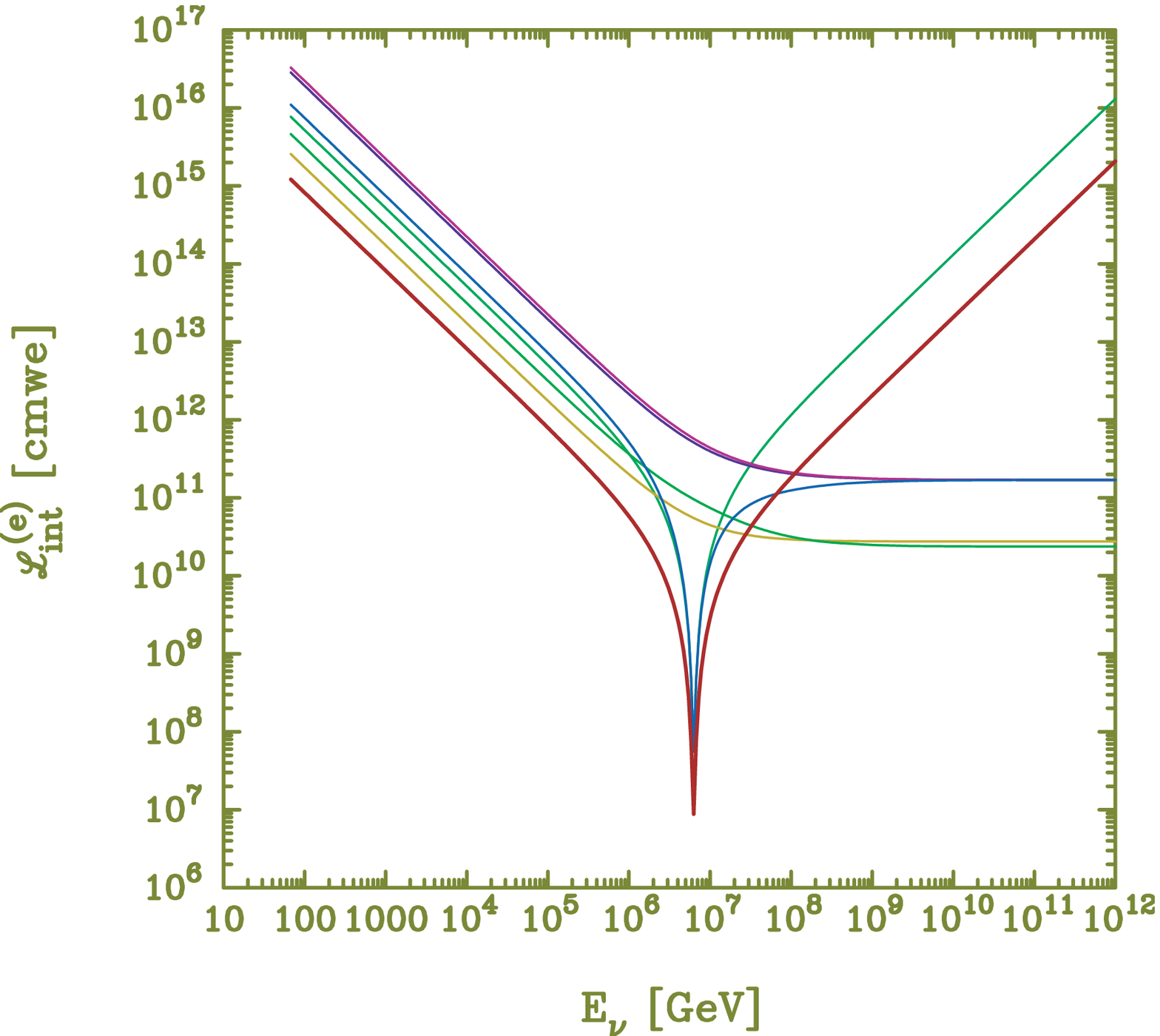}}
\caption{Left panel: Interaction length $\mathcal{L}_{\mathrm{int}}^{\nu 
N} = 1/\sigma_{\nu N}(E_{\nu})N_{\mathrm{A}}$, where $N_{\mathrm{A}}$ 
is Avogadro's number, for the reactions $\nu N \to
\hbox{anything}$ as a function of the incident neutrino energy.  The
left-hand scale, in cmwe, is appropriate for terrestrial applications;
the right-hand scale, in Mpc for the current Universe, is appropriate
for transport over cosmological distances.{\protect 
\cite{Gandhi:1998ri,Barenboim:2004di}} 
 Right panel: Interaction lengths for
neutrino interactions on electron targets.  At low energies, from
smallest to largest interaction length, the processes are (i)
$\bar{\nu}_{e}e \rightarrow \hbox{ hadrons}$, (ii) $\nu_{\mu}e
\rightarrow \mu\nu_{e}$, (iii) $\nu_{e}e \rightarrow \nu_{e}e$, (iv)
$\bar{\nu}_{e}e \rightarrow \bar{\nu}_{\mu}\mu$, (v) $\bar{\nu}_{e}e
\rightarrow \bar{\nu}_{e}e$, (vi) $\nu_{\mu}e \rightarrow \nu_{\mu}e$,
(vii) $\bar{\nu}_{\mu}e \rightarrow
\bar{\nu}_{\mu}e$.{\protect\cite{Gandhi:1995tf}}}
\label{fig:lint}
\end{figure}

The rising cross sections have important implications for neutrino 
telescopes. The left panel of Figure~\ref{fig:lint} shows that the 
Earth is opaque to neutrinos with energies above $40\tev$. This means 
that the strategy of looking down to distinguish charged-current 
interactions from the rain of cosmic-ray muons needs to be 
modified at high energies. On the other hand, the Universe at large is 
exceptionally poor in nucleons, and so the $(\nu N)$ interaction length of 
ultrahigh-energy neutrinos in the cosmos is effectively infinite. The 
right panel of Figure~\ref{fig:lint} shows the interaction cross 
section for neutrinos on electrons in the Earth, which is generally 
several orders of magnitude longer than the $\nu N$ interaction 
length. An important exception is the $\bar{\nu}_{e}e \to W^{-}$ 
resonance at $E_{\nu} \approx 6 \times 10^{15}\ev$.

\section{New Physics in Neutrino-Nucleon Interactions?}
New physics typically contributes equally to charged-current and
neutral-current cross sections, whereas standard electroweak
interactions favor the charged-current over neutral current, by a
factor of two or three.  A step or bump in the
neutral-current to charged-current ratio, measured as a function of
energy, is thus an excellent diagnostic for the onset of new phenomena.  The
example of squark production through $R$-parity--violating interactions
that we studied some time ago\cite{Carena:1998gd} offers a specific
illustration of this general rule.

\section{Influence of Neutrino Oscillations}
In the early days of planning for neutrino telescopes, people noticed
that observing $\tau$ production through the double-bang signature
might provide evidence for neutrino oscillations, since---to good
approximation---no $\nu_{\tau}$ are produced in conventional sources of
ultrahigh-energy neutrinos.  The discovery of neutrino oscillations is
of course already made; the phenomenon of neutrino oscillations means
that the flavor mixture at Earth, $\Phi = \{\varphi_{e},\varphi_{\mu},
\varphi_{\tau}\}$, will be different from the source mixture $\Phi^{0}
= \{\varphi_{e}^{0}, \varphi_{\mu}^{0}, \varphi_{\tau}^{0}\}$. The 
essential fact is that the vacuum oscillation length is very short, 
in cosmic terms. For  $|\Delta m^2| = 10^{-5}\ev^2$, the oscillation 
length
\begin{eqnarray}
\mathcal{L}_{\mathrm{osc}} & = & 4\pi E_{\nu}/|\Delta m^2| 
  \approx  2.5 \times 10^{-24}\mpc\cdot (E_{\nu}/1\ev)
  \label{eq:osclength}
\end{eqnarray}
is a fraction of a megaparsec, even for $E_{\nu}=10^{20}\ev$.
{Accordingly,} neutrinos oscillate many times  between cosmic source 
and terrestrial detector.

Neutrinos in the flavor basis $\ket{\nu_{\alpha}}$ are connected to the
mass eigenstates $\ket{\nu_{i}}$ by a unitary mixing matrix,
$\ket{\nu_{\alpha}} = \sum_i {U_{\beta i}}\ket{\nu_i}$.  It is
convenient to idealize $\sin\theta_{13} = 0$, $\sin 2\theta_{23} = 
1$, and consider
\begin{equation}
U_{\mathrm{ideal}} = \pmatrix{ c_{12} & s_{12} & 0 \cr
-s_{12}/\sqrt{2} & c_{12}/\sqrt{2} & 1/\sqrt{2} \cr
s_{12}/\sqrt{2} & -c_{12}/\sqrt{2} & 1/\sqrt{2}} \; .
\end{equation}
Then the transfer matrix $\mathcal{X}$ that maps the source flux 
$\Phi^{0}$ into the flux at Earth $\Phi$ takes the form
\renewcommand{\arraystretch}{1.5}
\begin{equation}
    \mathcal{X}_{\mathrm{ideal}} = \left(  
    \begin{array}{ccc}
    1-2x & x & x \\
    x & \textstyle{\frac{1}{2}}(1-x) & \textstyle{\frac{1}{2}}(1-x) \\
    x & \textstyle{\frac{1}{2}}(1-x) & \textstyle{\frac{1}{2}}(1-x)
    \end{array}\right)\;,
    \label{idealtrans}
\end{equation}
where $x = \sin^{2}\theta_{12}\cos^{2}\theta_{12}$.  Because the second
and third rows are identical, the $\nu_{\mu}$ and $\nu_{\tau}$ fluxes
that result from any source mixture $\Phi^{0}$ are equal:
$\varphi_{\mu} = \varphi_{\tau}$.  Independent of the value of $x$,
$\mathcal{X}_{\mathrm{ideal}}$ maps $\Phi^0_{\mathrm{std}} \rightarrow
\{\cfrac{1}{3},\cfrac{1}{3},\cfrac{1}{3}\}$.\footnote{I owe this 
formulation to Stephen Parke.}

\begin{figure}[tb]
    \centerline{\includegraphics[width=7.5cm]{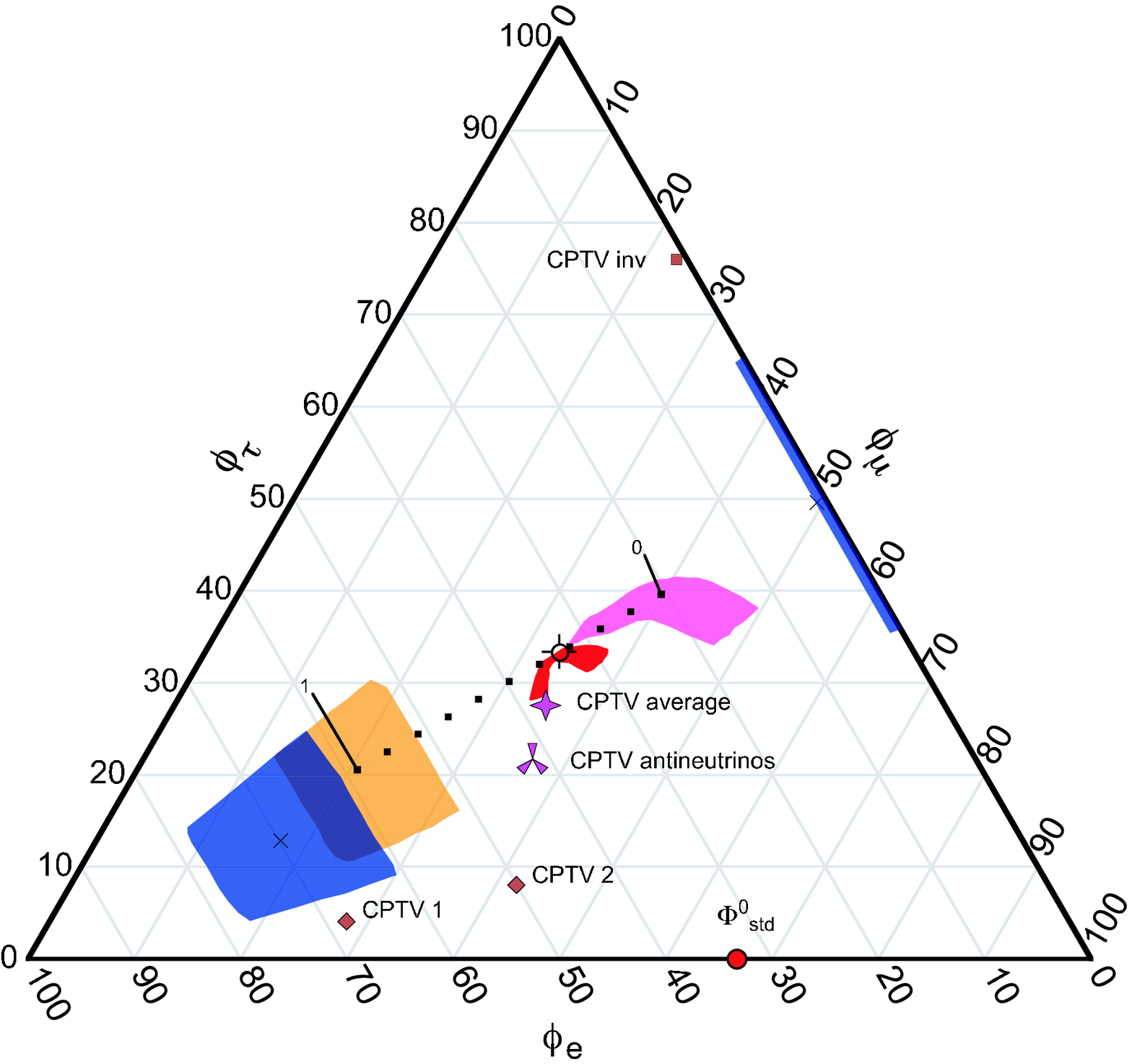}\quad
    \includegraphics[width=7.5cm]{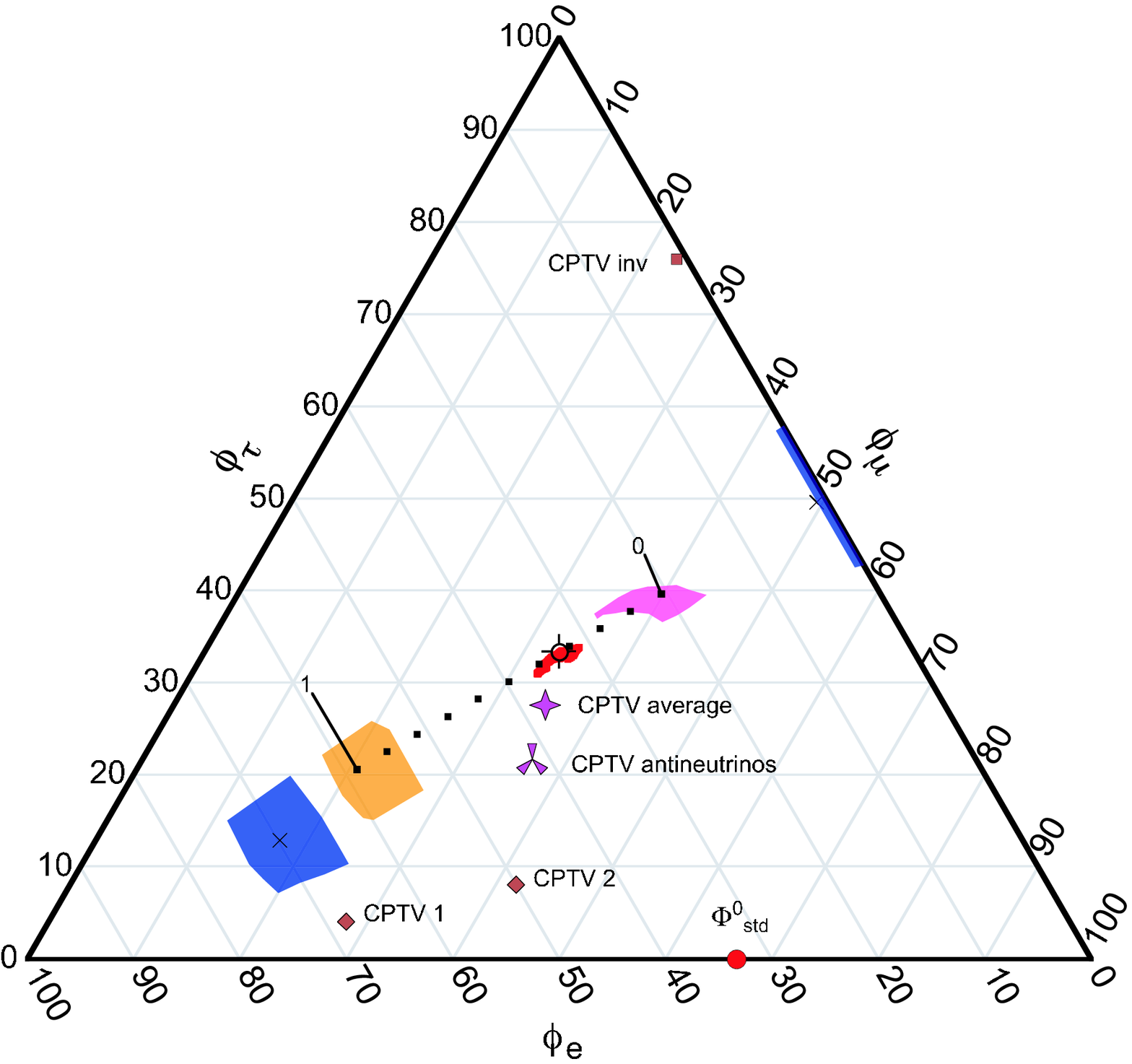}}
\caption{Ternary plots of the neutrino flux $\Phi$ at Earth, showing
the implications of current (left pane) and future (right pane)
knowledge of neutrino mixing.  The small black squares indicate the
$\nu_{e}$ fractions produced by the idealized transfer matrix
$\mathcal{X}_{\mathrm{ideal}}$ as $\varphi_{e}^{0}$ varies from 0 to 1
in steps of 0.1.  A crossed circle marks the standard mixed spectrum at
Earth, $\Phi_{\mathrm{std}} =
\{\cfrac{1}{3},\cfrac{1}{3},\cfrac{1}{3}\}$; for comparison, a red dot
marks the standard source spectrum, $\Phi_{\mathrm{std}}^{0} =
\{\cfrac{1}{3},\cfrac{2}{3},0\}$.  Colored swaths delimit the fluxes at
Earth produced by neutrino oscillations from the source mixtures
$\Phi_0^0 = \{0,1,0\}$ (pink), $\Phi_{\mathrm{std}}^0$ (red), and
$\Phi_1^0 = \{1,0,0\}$ (orange), using 95\% CL ranges for the
oscillation parameters.  Black crosses ($\times$) show the mixtures at
Earth that follow from neutrino decay, assuming normal ($\varphi_e
\approx 0.7$) and inverted ($\varphi_e \approx 0$) mass hierarchies.
The blue bands show how current and future uncertainties blur the
predictions for neutrino decays.  The violet tripod indicates how
\textsf{CPT}-violating oscillations shape the mix of antineutrinos that
originate in a standard source mixture, and the violet cross averages
that $\bar{\nu}$ mixture with the standard neutrino mixture.  The brown
squares denote consequences of \textsf{CPT} violation for antineutrino
decays.\protect\cite{Barenboim:2003jm}}
\label{fig:oscnow}
\end{figure}

The variation of $\varphi_{e}$ with the $\nu_{e}$ source fraction
$\varphi_{e}^{0}$ is shown as a sequence of small black squares (for
$\varphi_e^0 = 0, 0.1, \ldots , 1$) in Figure~\ref{fig:oscnow} for the
value $x = 0.21$, which corresponds to $\theta_{12} = 0.57$, the
central value in a recent global analysis.\cite{deHolanda:2002iv}  The $\nu_{e}$
fraction at Earth ranges from $0.21$, for $\varphi_{e}^{0} = 0$, to
$0.59$, for $\varphi_{e}^{0} = 1$.
\renewcommand{\arraystretch}{1}

The simple analysis based on $\mathcal{X}_{\mathrm{ideal}}$ is useful
for orientation, but it is important to explore the range of
expectations implied by global fits to neutrino-mixing parameters.  We
take\cite{Barenboim:2003jm} $0.49 < \theta_{12} < 0.67$,
$\cfrac{\pi}{4}\times 0.8 < \theta_{23} < \cfrac{\pi}{4} \times 1.2$,
and $0 < \theta_{13} < 0.1$.  With current uncertainties in the
oscillation parameters, a standard source spectrum,
$\Phi_{\mathrm{std}}^{0} = \{\cfrac{1}{3},\cfrac{2}{3},0\}$, is mapped
by oscillations onto the red boomerang near $\Phi_{\mathrm{std}} =
\{\cfrac{1}{3},\cfrac{1}{3},\cfrac{1}{3}\}$ in the left pane of
Figure~\ref{fig:oscnow}.  Given that $\mathcal{X}_{\mathrm{ideal}}$
maps $\Phi_{\mathrm{std}}^{0} \rightarrow \Phi_{\mathrm{std}}$ for any
value of $\theta_{12}$, it does not come as a great surprise that the
target region is of limited extent. The variation of $\theta_{23}$ away
from $\cfrac{\pi}{4}$ breaks the identity
$\varphi_{\mu}\equiv\varphi_{\tau}$ of the idealized analysis.

One goal of neutrino observatories will be to characterize cosmic
sources by determining the source mix of neutrino flavors.  It is
therefore of interest to examine the outcome of different assumptions
about the source.  We show in the left pane of Figure~\ref{fig:oscnow}
the mixtures at Earth implied by current knowledge of the oscillation
parameters for source fluxes $\Phi_0^0 = \{0,1,0\}$ (the purple band
near $\varphi_e \approx 0.2$) and $\Phi_1^0 = \{1,0,0\}$ (the orange
band near $\varphi_e \approx 0.6$).
For the $\Phi^0_{\mathrm{std}}$ and $\Phi^0_1$ source spectra, the
uncertainty in $\theta_{12}$ is reflected mainly in the variation of
$\varphi_e$, whereas the uncertainty in $\theta_{23}$ is expressed in
the variation of $\varphi_{\mu}/\varphi_{\tau}$ For the $\Phi_0^0$
case, the influence of the two angles is not so orthogonal.  For all
the source spectra we consider, the uncertainty in $\theta_{13}$ has
little effect on the flux at Earth.  The extent of the three regions,
and the absence of a clean separation between the regions reached from
$\Phi_{\mathrm{std}}^{0}$ and $\Phi_0^0$ indicates that characterizing
the source flux will be challenging, in view of the current
uncertainties of the oscillation parameters.

Over the next five years---roughly the time scale on which large-volume
neutrino telescopes will come into operation---we can anticipate
improved information on $\theta_{12}$ and $\theta_{23}$ from KamLAND
and the long-baseline accelerator experiments at
Soudan  and Gran Sasso.  We base our
projections for the future on the ranges $0.54 < \theta_{12} < 0.63$
and $\cfrac{\pi}{4} \times 0.9 < \theta_{23} < \cfrac{\pi}{4} \times
1.1$, still with $0 < \theta_{13} < 0.1$. The results are shown in the
right panel of Figure~\ref{fig:oscnow}.  The (purple) target region for
the source flux $\Phi_0^0$ shrinks appreciably and separates from the
(red) region populated by $\Phi_{\mathrm{std}}^0$, which is now tightly
confined around $\Phi_{\mathrm{std}}$.  The (orange) region mapped from
the source flux $\Phi_1^0$ by oscillations shrinks by about a factor of
two in the $\varphi_e$ and $\varphi_{\mu}-\varphi_{\tau}$ dimensions.


\section{Reconstructing the Neutrino Mixture at the Source\cite{Barenboim:2003jm}}
\begin{figure}[tb]
    \centerline{\includegraphics[width=10cm]{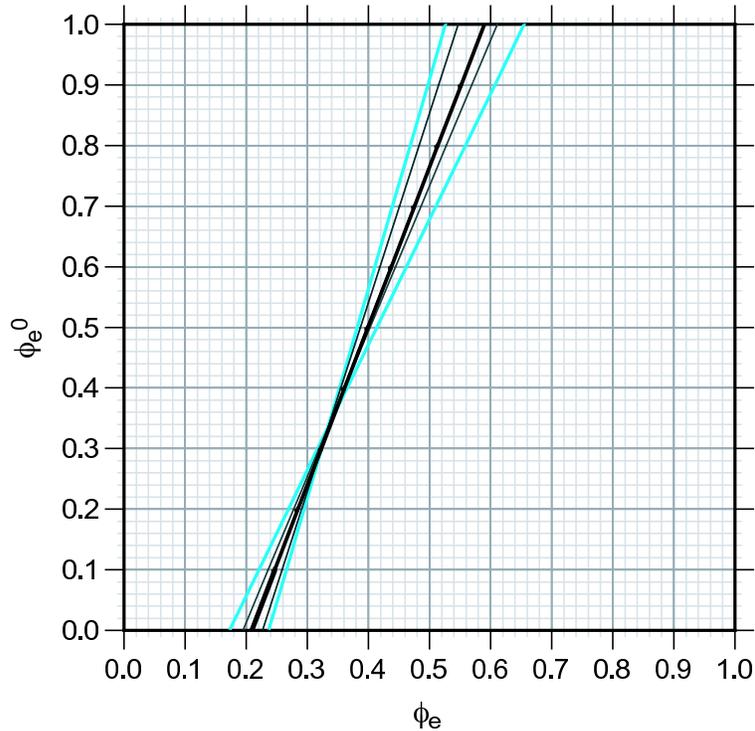}}
    \caption{The source flux $\varphi_e^0$ of electron neutrinos
    reconstructed from the $\nu_e$ flux $\varphi_e$ at Earth, using the
    ideal transfer matrix $\mathcal{X}_{\mathrm{ideal}}$ of Eqn.\
    (\ref{idealtrans}).  The heavy solid line refers to 
    $\theta_{12} = 0.57$.  The light blue lines refer to
    the current experimental constraints (at 95\% CL), and the thin
    solid lines refer to a projection of future experimental
    constraints.\protect\cite{Barenboim:2003jm}}
\label{fig:invert}
\end{figure}

What can observations of the blend $\Phi$ of neutrinos arriving at
Earth tell us about the source?  Inferring the nature of the processes
that generate cosmic neutrinos is more complicated than it would be if 
neutrinos did not oscillate.  Because $\nu_{\mu}$ and
$\nu_{\tau}$ are fully mixed---and thus enter identically in
$\mathcal{X}_{\mathrm{ideal}}$---it is not possible fully to characterize
$\Phi^{0}$. We can, however, reconstruct the $\nu_{e}$ fraction at the 
source as $\varphi_{e}^{0} = (\varphi_{e} - x)/(1 - 3x)$, where $x = 
\sin^{2}\theta_{12}\cos^{2}\theta_{12}$. 
The reconstructed source flux $\varphi_e^0$ is shown in
Figure~\ref{fig:invert} as a function of the $\nu_e$ flux at Earth.  The
heavy solid line represents the best-fit value for $\theta_{12}$; the
light blue lines and thin solid lines indicate the current and future
95\% CL bounds on $\theta_{12}$.

A possible strategy for beginning to characterize a source of cosmic
neutrinos might proceed by measuring the $\nu_e/\nu_{\mu}$ ratio and
estimating $\varphi_e$ under the plausible assumption---later to be
checked---that $\varphi_{\mu} = \varphi_{\tau}$.  Very large
($\varphi_e \gtap 0.65$) or very small ($\varphi_e \ltap 0.15$) $\nu_e$
fluxes cannot be accommodated in the standard neutrino-oscillation
picture.  Observing an extreme $\nu_e$ fraction would implicate
unconventional physics.  

Determining the energy dependence of $\varphi_{e}^{0}$ may also be of
astrophysical interest.  In a thick source, the highest energy muons
may interact and lose energy before they can decay.  In the limit of
$\varphi_{e}^{0} = 0$, the arriving flux will be $\Phi =
\{x,\cfrac{1}{2}(1-x),\cfrac{1}{2}(1-x)\} \approx \{\cfrac{1}{5},
\cfrac{2}{5}, \cfrac{2}{5}\}$ (cf.\ Figure~\ref{fig:oscnow}).  More
generally, measured $\nu_e$ fractions that depart significantly from
the canonical $\varphi_e = \cfrac{1}{3}$ would suggest nonstandard
neutrino sources.  An observed flux $\varphi_e = 0.5 \pm 0.1$ points to
a source flux $0.47 \ltap \varphi_e^0 \ltap 1$, with current
uncertainties, whereas $\varphi_e = 0.25 \pm 0.10$ indicates $0 \ltap
\varphi_e^0 \ltap 0.35$.

Constraining the source flux sufficiently to test the nature of the
neutrino production process will require rather precise determinations
of the neutrino flux at Earth.  Suppose we want to test the idea that
the source flux has the standard composition $\Phi^{0}_{\mathrm{std}}$.
With today's uncertainty on $\theta_{12}$, a 30\% measurement that
locates $\varphi_e = 0.33 \pm 0.10$ implies only that $0 \ltap
\varphi_e^0 \ltap 0.68$.  For a measured flux in the neighborhood of
$\cfrac{1}{3}$, the uncertainty in the solar mixing angle is of little
consequence: the constraint that arises if we assume the central value
of $\theta_{12}$ is not markedly better: $0.06 \ltap \varphi_e^0 \ltap
0.59$.  A 10\% measurement of the $\nu_e$ fraction, $\varphi_e = 0.33
\pm 0.033$, would make possible a rather restrictive constraint on the
nature of the source.  The central value for $\theta_{12}$ leads to
$0.26 \ltap \varphi_e^0 \ltap 0.43$, blurred to $0.22 \ltap \varphi_e^0
\ltap 0.45$ with current uncertainties.

\section{Influence of Neutrino Decays}
Beacom and Bell~\cite{Beacom:2002cb} have shown that observations of
solar neutrinos set the most stringent plausible lower bound on the
reduced lifetime of a neutrino of mass $m$ as $\tau/m \gtap
10^{-4}\hbox{ s/eV}$.  This rather modest limit opens the possibility
that some neutrinos do not survive the journey from astrophysical
sources.  Decays of unstable neutrinos over cosmic distances can lead
to mixtures at Earth that are incompatible with the oscillations of
stable neutrinos.\cite{Beacom:2002cb,Beacom:2002vi,Barenboim:2003jm}
The candidate decays are transitions between mass eigenstates by
emission of a very light particle, $\nu_i \rightarrow (\nu_j,
\bar{\nu}_j)+X$.  Dramatic effects occur when the decaying neutrinos
disappear, either by decay to invisible products or by decay into
active neutrino species so degraded in energy that they contribute
negligibly to the total flux at the lower energy.

If the lifetimes of the unstable mass eigenstates are short compared
with the flight time from source to Earth, all the unstable neutrinos 
will decay, and the (unnormalized) flavor
$\nu_{\alpha}$ flux at Earth will be $\widetilde{\varphi}_{\alpha} (E_{\nu})=  
\sum_{i = \mathrm{stable} } \sum_{\beta} \varphi^{0}_{\beta}(E_{\nu})
|U_{\beta i}|^2 |U_{\alpha i}|^2$,
with $\varphi_{\alpha} = \widetilde{\varphi}_{\alpha} / \sum_{\beta}
\widetilde{\varphi}_{\beta}$.
Should only the lightest neutrino survive, the flavor mix of neutrinos
arriving at Earth is determined by the flavor composition of the
lightest mass eigenstate, \textit{independent of the flavor mix at the
source.} 

For a normal mass hierarchy $m_1 < m_2 < m_3$, the $\nu_{\alpha}$ flux
at Earth is $\varphi_{\alpha} = |U_{\alpha 1}|^2$.  {Consequently,} the
neutrino flux at Earth is $\Phi_{\mathrm{normal}} = \{|U_{e1}|^2,
|U_{\mu1}|^2,|U_{\tau1}|^2\} \approx \{0.70, 0.17, 0.13\}$ for our
chosen central values of the mixing angles.  If the mass hierarchy is
inverted, $m_1 > m_2 > m_3$, the lightest (hence, stable) neutrino is
$\nu_3$, so the flavor mix at Earth is determined by $\varphi_{\alpha}
= |U_{\alpha 3}|^2$.  In this case, the neutrino flux at Earth is
$\Phi_{\mathrm{inverted}} = \{|U_{e3}|^2, |U_{\mu3}|^2,|U_{\tau3}|^2\}
\approx \{0, 0.5, 0.5\}$.  Both $\Phi_{\mathrm{normal}}$ and
$\Phi_{\mathrm{inverted}}$, which are indicated by crosses ($\times$)
in Figure~\ref{fig:oscnow}, are very different from the standard flux
$\Phi_{\mathrm{std}} = \{\varphi_{e} = \cfrac{1}{3}, \varphi_{\mu} =
\cfrac{1}{3}, \varphi_{\tau} = \cfrac{1}{3}\}$ produced by the ideal
transfer matrix from a standard source.  Observing either mixture would
represent a departure from conventional expectations.

The fluxes that result from neutrino decays \textit{en route} from the
sources to Earth are subject to uncertainties in the neutrino-mixing
matrix.  The expectations for the two decay scenarios are indicated by
the blue regions in Figure~\ref{fig:oscnow}, where we indicate the
consequences of 95\% CL ranges of the mixing parameters now and in the
future.  With current uncertainties, the normal hierarchy populates
$0.61 \ltap \varphi_e \ltap 0.77$, and allows considerable departures
from $\varphi_{\mu} = \varphi_{\tau}$.  The normal-hierarchy decay
region based on current knowledge overlaps the flavor mixtures that
oscillations produce in a pure-$\nu_e$ source, shown in orange.  (It
is, however, far removed from the standard region that encompasses
$\Phi_{\mathrm{std}}$.)  With the projected tighter constraints on the
mixing angles, the range in $\varphi_e$ swept out by oscillation from a
pure-$\nu_e$ source or decay from a normal hierarchy shrinks by about a
factor of two.  Neutrino decay then populates $0.65 \ltap \varphi_e 
\ltap 0.74$, and is separated from the oscillations.  The degree of
separation between the region populated by normal-hierarchy decay and
the one populated by mixing from a pure-$\nu_e$ source depends on the
value of the solar mixing angle $\theta_{12}$.  For the seemingly
unlikely value $\theta_{12} = \cfrac{\pi}{4}$, both mechanisms yield
$\Phi = \{\cfrac{1}{2}, \cfrac{1}{4}, \cfrac{1}{4}\}$.

The mixtures that result from the decay of the heavier members of an
inverted hierarchy entail $\varphi_e\approx 0$.  These mixtures are
well separated from any that would result from neutrino oscillations,
for any conceivable source at cosmic distances.

The energies of neutrinos that may be detected in the future from AGNs
and other cosmic sources range over several orders of magnitude,
whereas the distances to such sources vary over perhaps one order of
magnitude.  The neutrino energy sets the neutrino lifetime in the
laboratory frame; more energetic neutrinos survive over longer flight
paths than their lower-energy companions.\footnote{A similar phenomenon
is familiar for cosmic-ray muons.}  Under propitious circumstances of
reduced lifetime, path length, and neutrino energy, it might be
possible to observe the transition from more energetic survivor
neutrinos to less energetic decayed neutrinos.

\begin{figure}
    \centerline{\includegraphics[width=7.5cm]{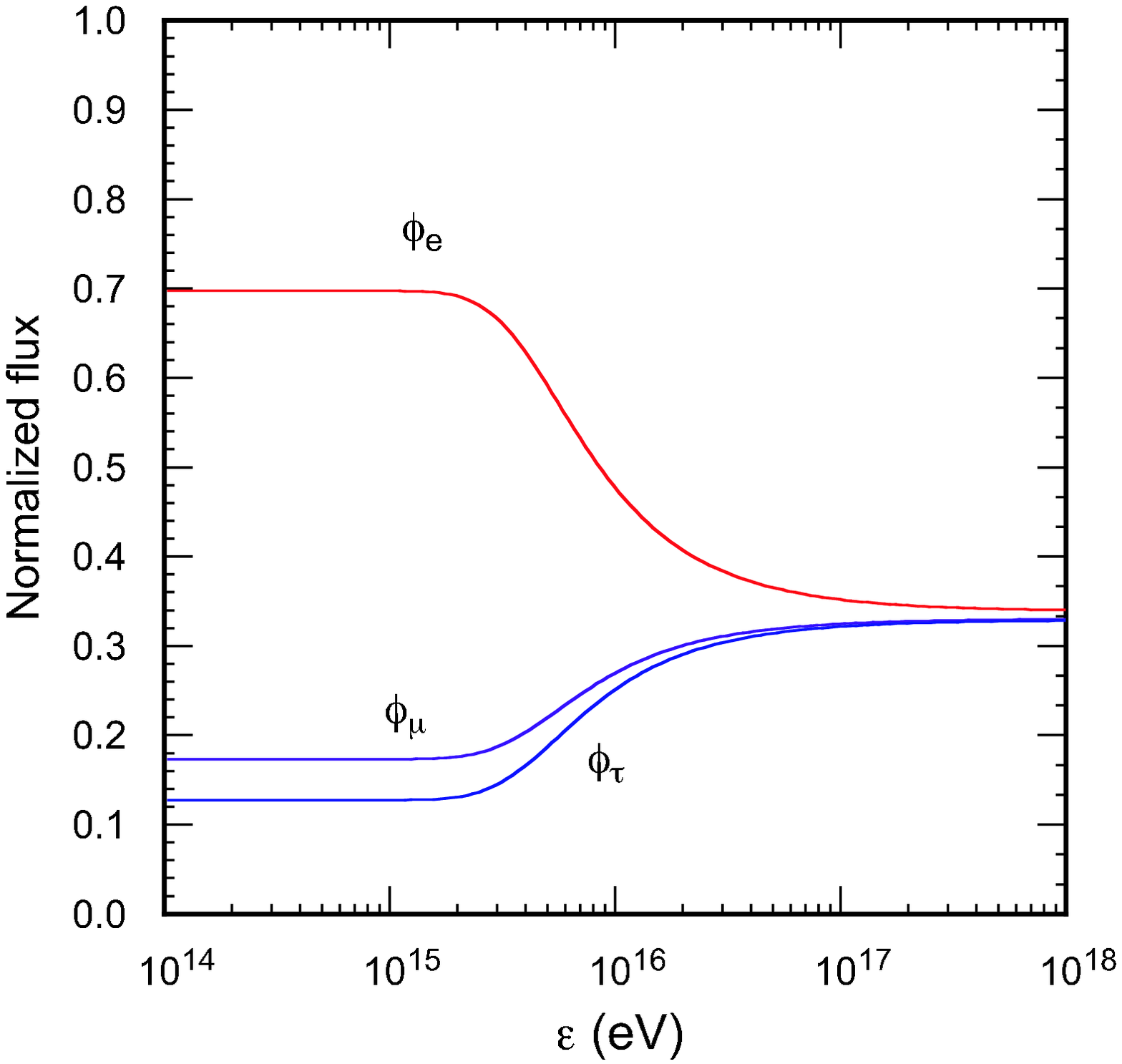}\quad
    \includegraphics[width=7.5cm]{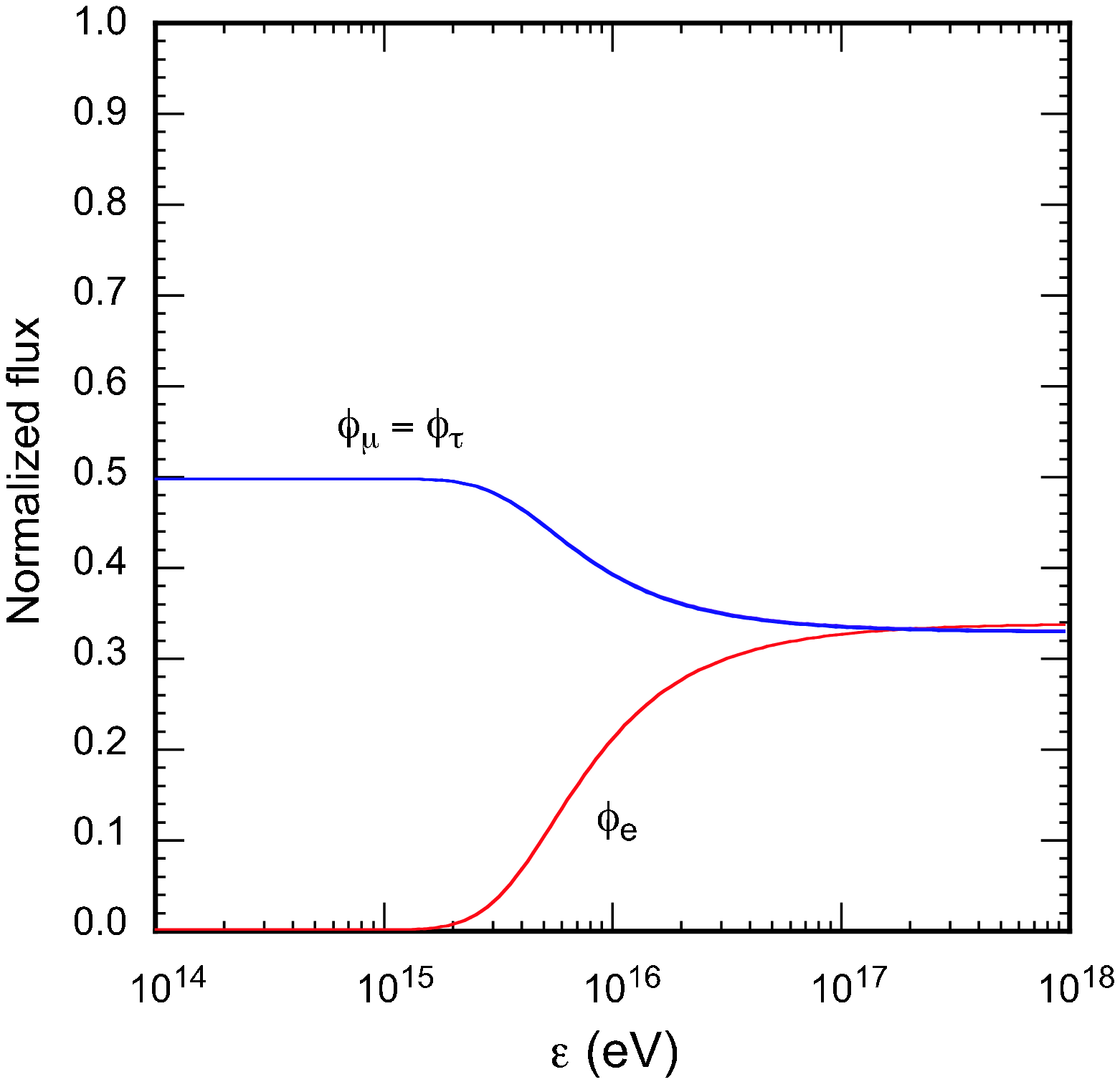}}
\caption{Energy dependence of normalized $\nu_e$, $\nu_{\mu}$, and
$\nu_{\tau}$ fluxes, for the two-body decay of the two upper mass
eigenstates, with the neutrino source at $L=100\hbox{ Mpc}$ from Earth
and $\tau/m = 1\hbox{ s/eV}$.  The left panel shows the result for a
normal mass hierarchy; the right panel shows the result for an inverted
mass hierarchy.  With suitable rescaling of the neutrino energy 
[$E_{\nu} = \varepsilon (1\s/\hbox{eV})/(\tau_{\nu}/m_{\nu})\cdot
L/(100\mpc)$], these plots apply for any combination of path length and
reduced lifetime.\protect\cite{Barenboim:2003jm}}
\label{fig:decays}
\end{figure}

If decay is not complete, the (unnormalized) flavor $\nu_{\alpha}$ flux
arriving at Earth from a source at distance $L$ is given by
$\widetilde{\varphi}_{\alpha} (E_{\nu})=  
\sum_{i } \sum_{\beta} \varphi^{0}_{\beta}(E_{\nu})
|U_{\beta i}|^2 |U_{\alpha i}|^2 e^{-(L/E_{\nu})(m_i/\tau_i)}$,
with the normalized flux $\varphi_{\alpha}(E_{\nu}) =
\widetilde{\varphi}_{\alpha}(E_{\nu}) / \sum_{\beta}
\widetilde{\varphi}_{\beta}(E_{\nu})$.  An idealized case will
illustrate the possibilities for observing the onset of neutrino decay
and estimating the reduced lifetime.  Assume a normal mass hierarchy,
$m_1 < m_2 < m_3$, and let $\tau_3/m_3 = \tau_2/m_2 \equiv \tau/m$.
For a given path length $L$, the neutrino energy at which the
transition occurs from negligible decays to complete decays is
determined by $\tau/m$.  The left pane of
Figure~\ref{fig:decays} shows the energy evolution of the normalized neutrino
fluxes arriving from a standard source; the energy scale is appropriate
for the case $\tau/m = 1\hbox{ s/eV}$ and $L = 100\hbox{ Mpc}$.  

If we locate the transition from survivors to decays at neutrino energy
$E^{\star}$, then we can estimate the reduced lifetime in terms of the
distance to the source as
\begin{equation}
\tau / m \approx 100\hbox{ s/eV} \cdot \left(\frac{L}{\hbox{1
Mpc}}\right) \left( \frac{1\hbox{ TeV}}{E^{\star}}\right)\; .
\label{taumest}
\end{equation}
In practice, ultrahigh-energy neutrinos are likely to arrive from a
multitude of sources at different distances from Earth, so the
transition region will be blurred.\footnote{Our assumption that
$\tau_3/m_3 = \tau_2/m_2 \equiv \tau/m$ is also a special case.}
Nevertheless, it would be rewarding to observe the decay-to-survival
transition, and to use Eqn.\ (\ref{taumest}) to estimate---even within
one or two orders of magnitude---the reduced lifetime.  If no evidence
appears for a flavor mix characteristic of neutrino decay, then Eqn.\
(\ref{taumest}) provides a lower bound on the neutrino lifetime.  For
that purpose, the advantage falls to large values of $L/E^{\star}$, and
so to the lowest energies at which neutrinos from distant sources can
be observed.  Observing the standard flux, $\Phi_{\mathrm{std}} =
\{\cfrac{1}{3},\cfrac{1}{3},\cfrac{1}{3}\}$, which is incompatible with
neutrino decay, would strengthen the current bound on $\tau/m$ by some
seven orders of magnitude, for 10-TeV neutrinos from sources at
$100\mpc$.

\section{UHE Neutrino Annihilation on Relic Neutrinos}
The neutrino gas that we believe permeates the present Universe has
never been detected directly.  By observing resonant annihilation
of extremely-high-energy cosmic neutrinos on the background neutrinos
through the reaction $\nu\bar{\nu} \to
Z^{0}$,\cite{Weiler:1982qy,Roulet:1992pz,Fargion:1997ft,Weiler:1997sh,Eberle:2004ua}
we could hope to confirm the presence of the relic neutrinos and 
learn the absolute neutrino masses and the flavor composition of the neutrino
mass eigenstates. I have recently made a detailed study of the 
prospects for cosmic-neutrino annihilation spectroscopy with Gabriela 
Barenboim and Olga Mena.\cite{Barenboim:2004di} I summarize some of 
our main findings here.

\begin{figure}
    \centerline{\includegraphics[width=10cm]{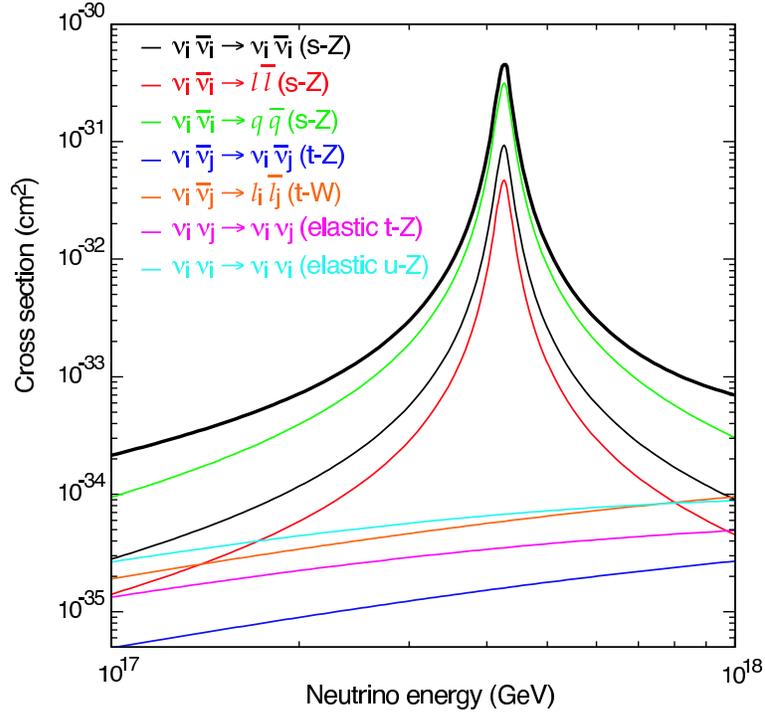}}
    \caption{Total neutrino annihilation cross section and the different
contributing channels as a function of the neutrino energy
assuming a relic neutrino mass of $m_\nu =10^{-5}\ev$ and zero
redshift.\protect\cite{Barenboim:2004di}}
\label{fig:zed}
\end{figure}

\begin{figure}
    \centerline{\includegraphics[width=8cm]{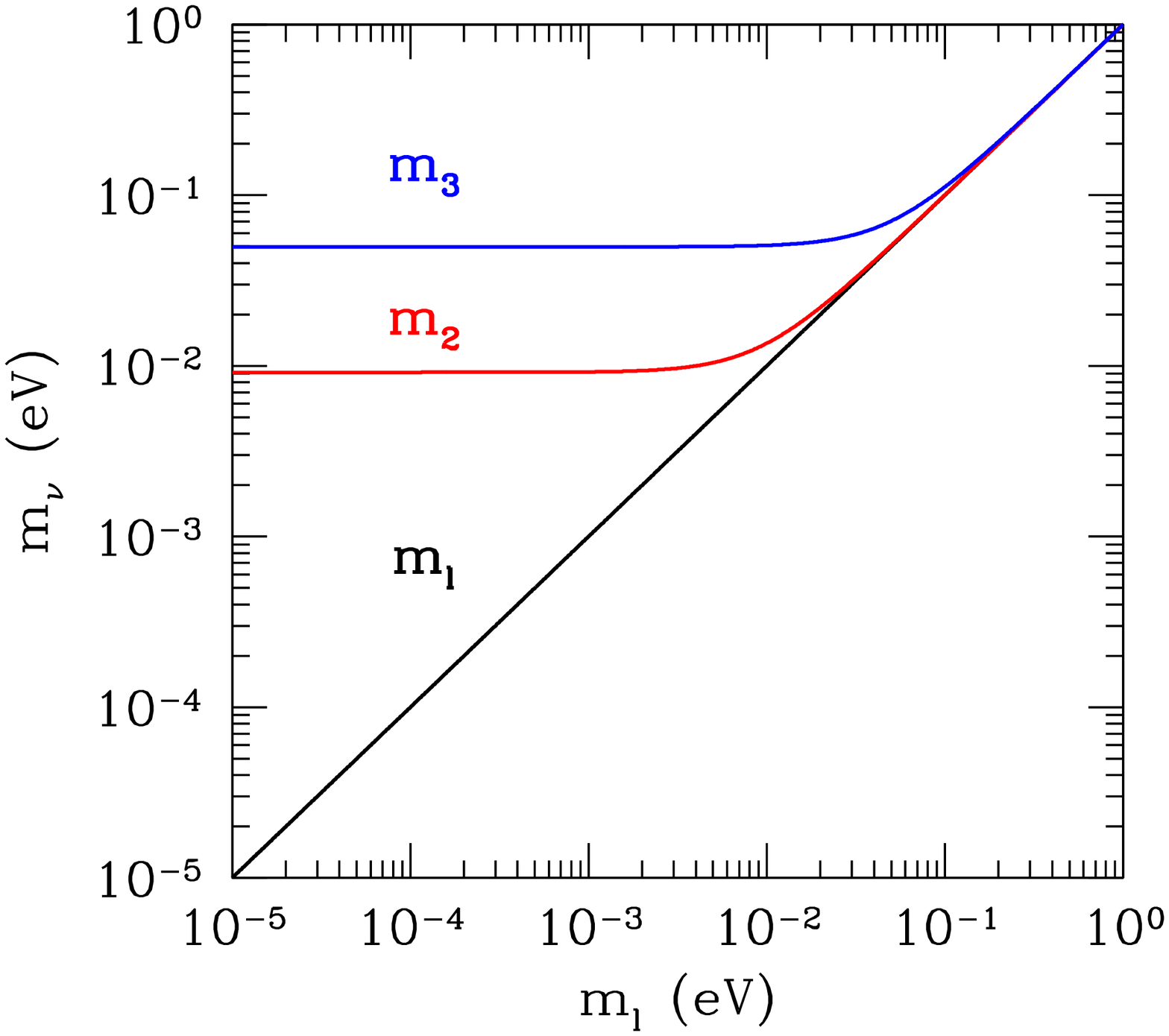}\quad
    \includegraphics[width=8cm]{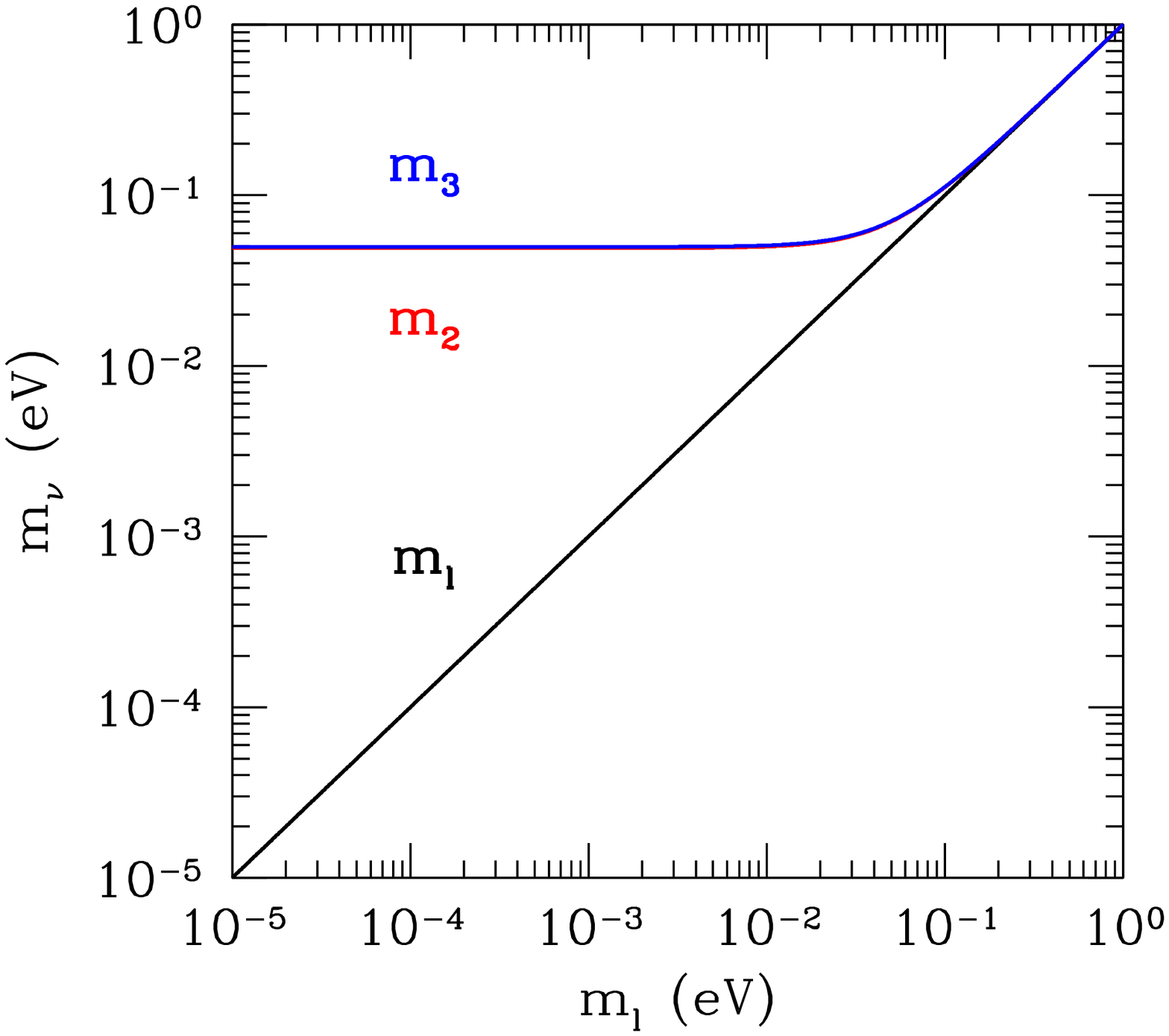}}
    \caption{Favored values for the neutrino masses as functions of the lightest 
 neutrino mass, $m_\ell$, in the three neutrino scenario for normal 
 hierarchy (left panel, $m_{\ell} = m_{1}$) and the inverted hierarchy
  (right panel, $m_{\ell} = m_{3}$).\protect\cite{Barenboim:2004di}}
\label{fig:neumasses}
\end{figure}

I present the components of the neutrino-(anti)neutrino cross sections 
in Figure~\ref{fig:zed}. 
The feature that matters is the $Z^{0}$-formation line that occurs 
near the resonant energy $E_{\nu}^{Z\mathrm{res}} = 
M_{Z}^{2}/2m_{\nu_{i}}$. Existing knowledge of neutrino oscillations 
allows us to project the neutrino mass spectrum in terms of the 
unknown mass of the lightest neutrino. The expectations are shown in  
Figure~\ref{fig:neumasses} for normal and inverted mass hierarchies.

In an idealized \textit{Gedankenexperiment,} we may consider an
extremely high-energy neutrino beam traversing a very long column with
the relic-neutrino properties of the current Universe.  We neglect for
now the expansion of the Universe and the thermal motion of the relic
neutrinos.  The ``cosmic neutrino attenuator'' is thus a column of
length $L$ with uniform neutrino density $n_{\nu0} = 56\cm^{-3}$ of
each neutrino species, $\nu_{e}, \bar{\nu}_{e}, \nu_{\mu},
\bar{\nu}_{\mu}, \nu_{\tau}, \bar{\nu}_{\tau}$.  If the column of relic
neutrinos is thick enough to attenuate neutrinos appreciably through
resonant absorption at the $Z^{0}$ gauge boson, the energies that
display absorption dips point to the neutrino masses through the
resonant-energy condition.  The relative depletion of
$\nu_{e},\nu_{\mu},\nu_{\tau}$ in each of the lines measures the flavor
composition of the corresponding neutrino mass eigenstate.

Even if we had at our disposal an adequate neutrino beam (with 
energies extending beyond $10^{26}\ev$), the time required
to traverse one interaction length for $\nu\bar{\nu} \to Z^{0}$
annihilation on the relic background in the current Universe ($1.2
\times 10^{4}\mpc = 39\hbox{ Gly}$) exceeds the age of the
Universe, not to mention the human attention span.  If we are ever to
detect the attenuation of neutrinos on the relic-neutrino background,
we shall have to make use of astrophysical or cosmological neutrinos
sources traversing the Universe over cosmic time scales.  The expansion
of the Universe over the propagation time of the neutrinos entails 
three important effects: the evolution of relic-neutrino density, the
redshift of the incident neutrino energy, and the redshift of the 
relic-neutrino temperature.

\begin{figure}
    \centerline{\includegraphics[width=8cm]{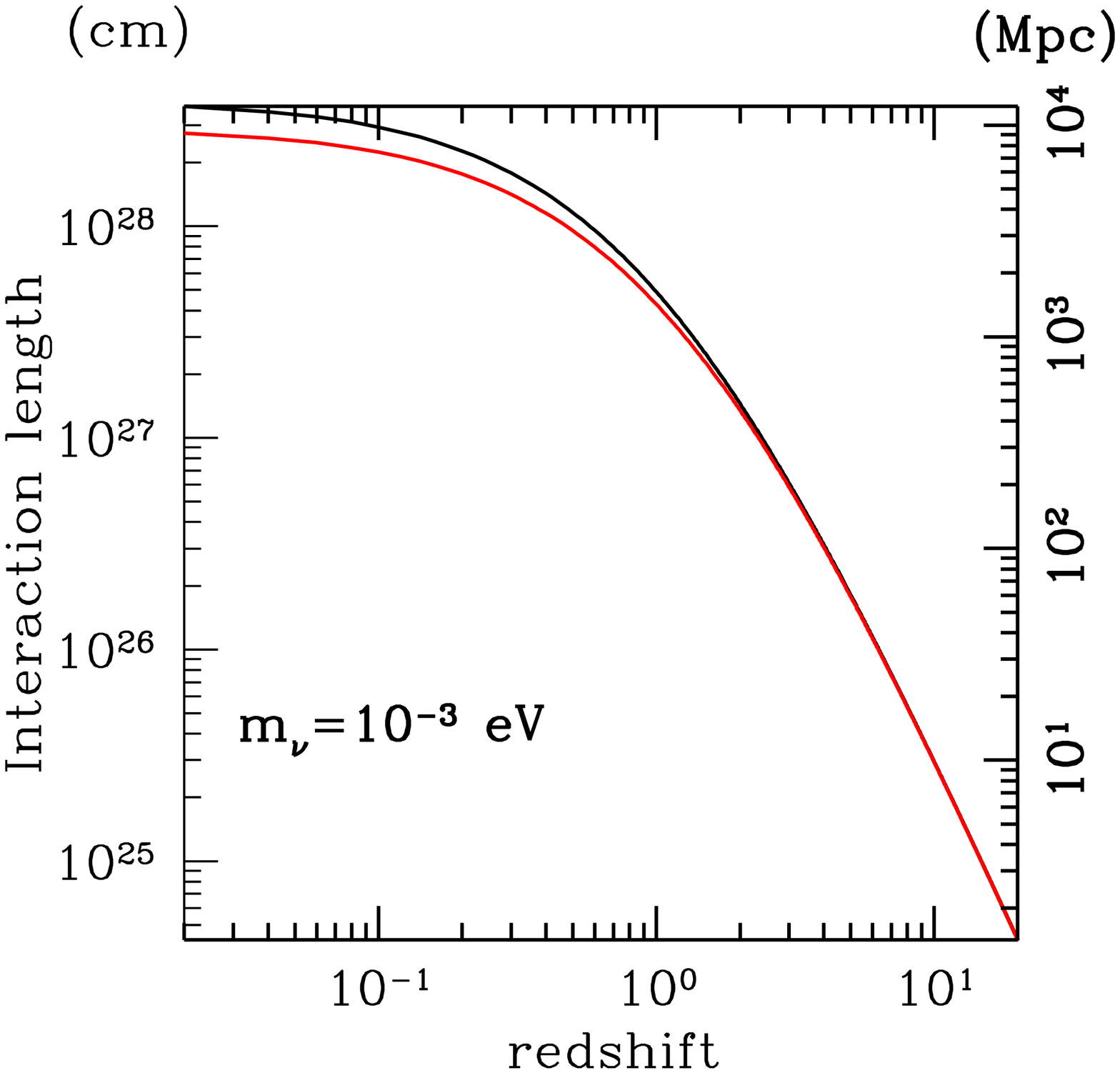}\quad
    \includegraphics[width=8cm]{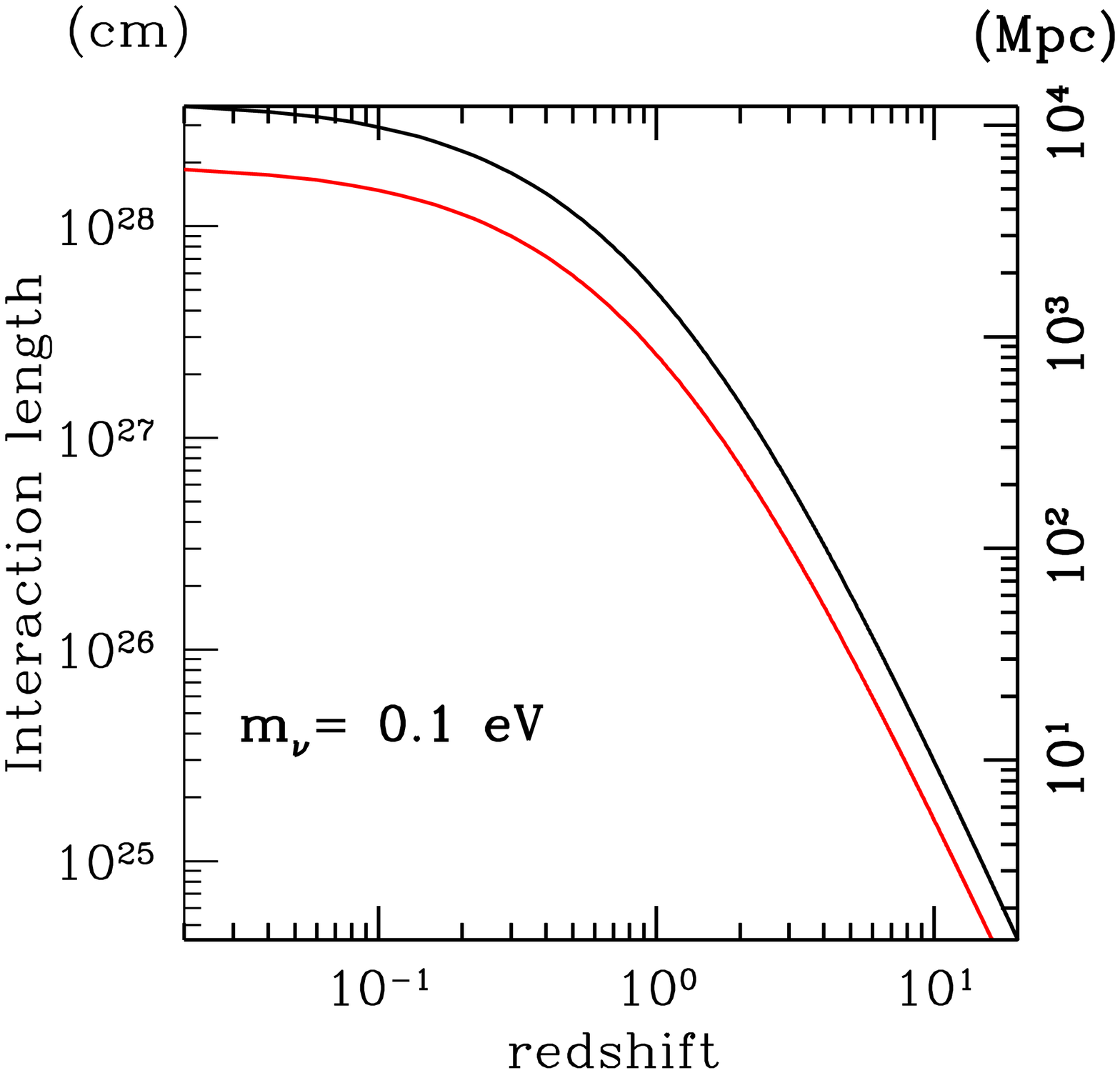}}
    \caption{Interaction lengths 
versus redshift at the $Z^{0}$ resonance 
for neutrino masses $m_\nu =10^{-3}, 10^{-1}\ev$ (left,  
and right panels). The left-hand scales are in centimeters, the 
right-hand scales in megaparsecs.
In the center and right panels, the upper (black) 
line is for the  Dirac-neutrino case; the  lower (red) line applies to Majorana 
neutrinos.\protect\cite{Barenboim:2004di}}
\label{fig:lintZ}
\end{figure}

\begin{figure}
    \centerline{\includegraphics[width=8cm]{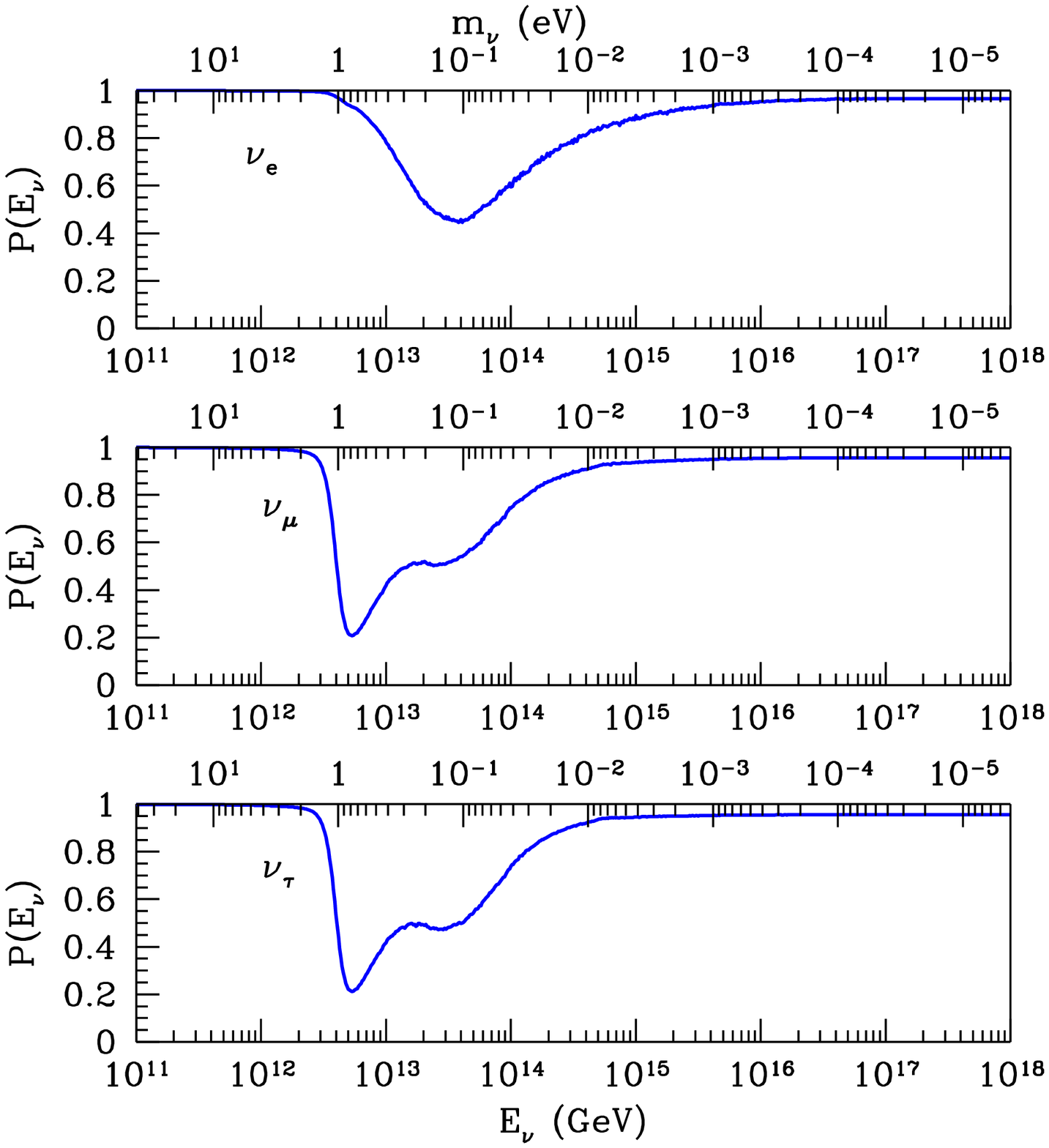}\quad
    \includegraphics[width=8cm]{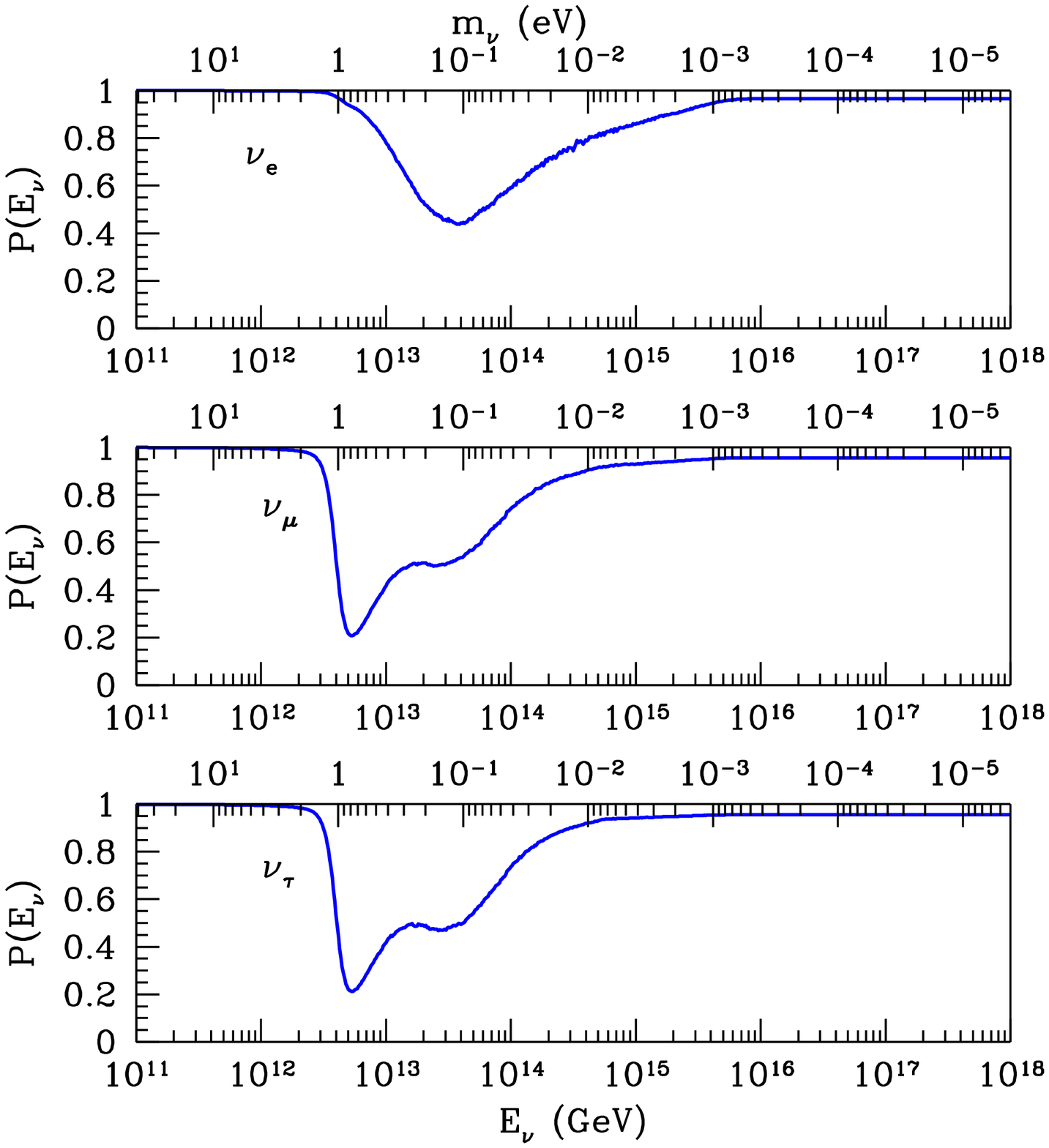}}
    \caption{Survival probabilities for $\nu_e$, $\nu_\mu$, and 
 $\nu_\tau$  as a function of the neutrino energy,
 after integration back to redshift $z = 20$, taking into account the 
 Fermi smearing induced by the thermal motion of the relic neutrinos.
The results apply for a normal hierarchy with lightest neutrino mass
 $m_{\ell}=10^{-5}\ev$ (left panel) or $m_{\ell}=10^{-3}\ev$ (right panel).
\protect\cite{Barenboim:2004di}}
\label{fig:fermimo}
\end{figure}

\begin{figure}
    \centerline{\includegraphics[width=8cm]{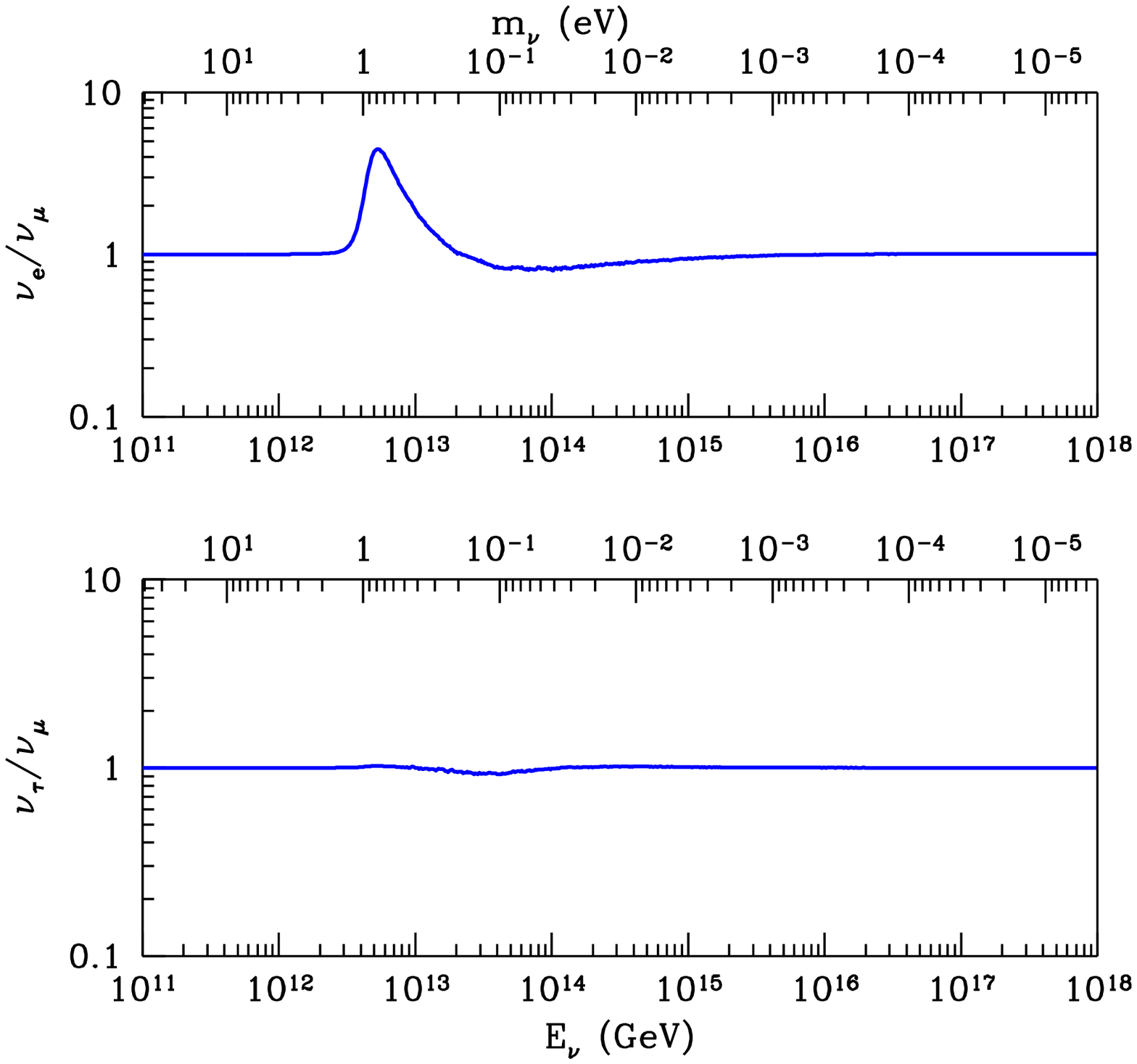}\quad
    \includegraphics[width=8cm]{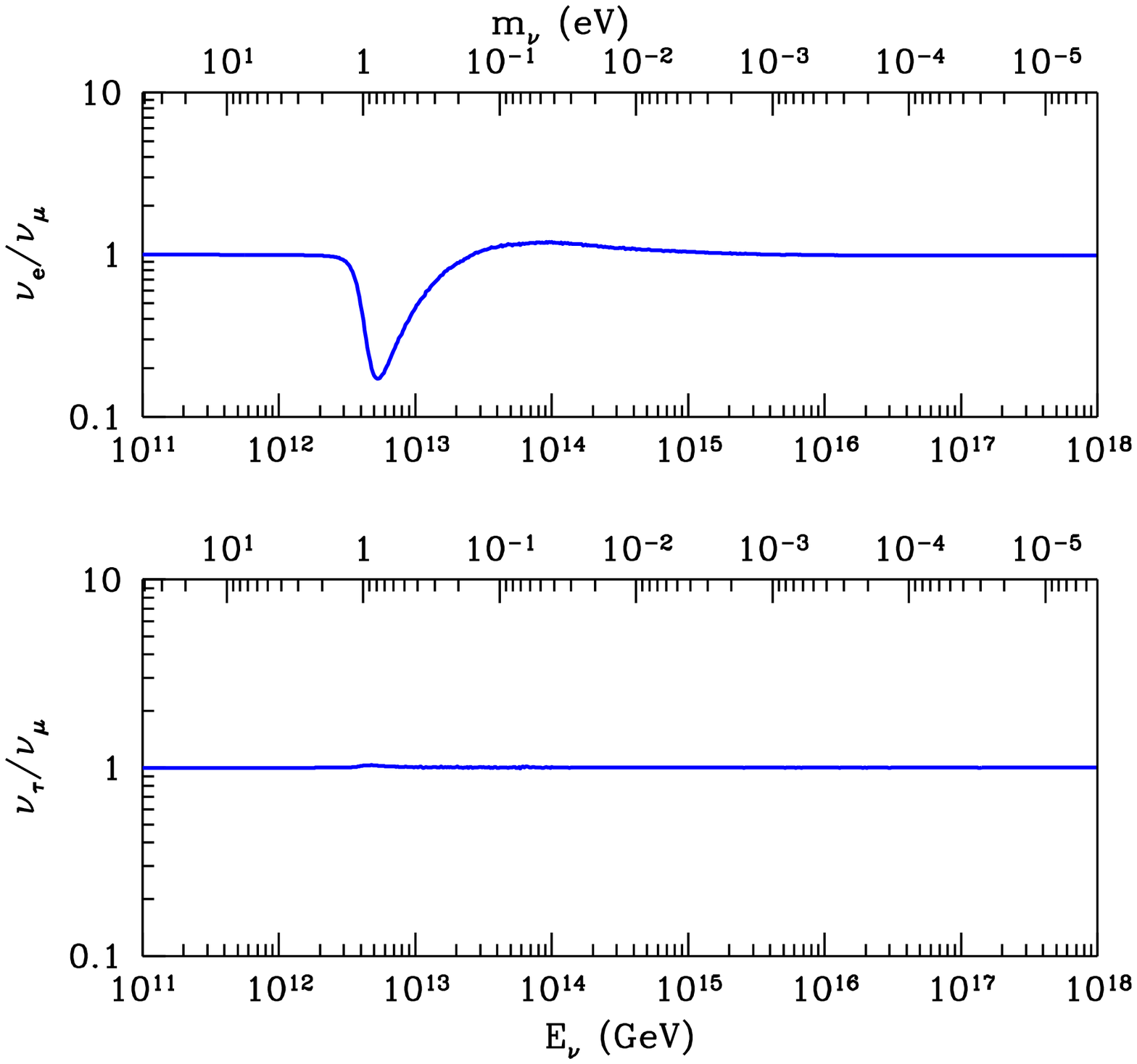}}
    \caption{Flux ratios $\nu_{e}/\nu_{\mu}$ and 
    $\nu_{\tau}/\nu_{\mu}$ at Earth, for normal (left panel) and 
    inverted (right panel) mass hierarchies with $m_{\ell} = 
    10^{-5}\ev$, after integration back to redshift $z=20$ and a 
    thermal averaging over the relic-neutrino momentum distribution.
    The scale at the top shows the neutrino mass 
    defined as $m_{\nu} = M_{Z}^{2}/2E_{\nu}$ that would be inferred if 
    neutrino energies were not redshifted.\protect\cite{Barenboim:2004di}}
\label{fig:fermimorats}
\end{figure}

The decrease of interaction lengths with increasing redshift shown in 
Figure~\ref{fig:lintZ} reveals that for redshifts in the range from 
one to ten, the interaction length matches the distance to the AGNs we 
consider as plausible UHE neutrino sources.\footnote{\ldots though not 
perhaps with the energies required here!}
The absorption lines that result from a full calculation, including 
the effects of the relics' Fermi motion and the evolution of the 
Universe back to redshift $z = 20$, are shown in  Figure~\ref{fig:fermimo} 
for two values of the lightest neutrino mass, $m_{\ell} = 10^{-5}$ 
and $10^{-3}\ev$. Although the lines are distorted and displaced from 
their natural shapes and positions by redshifting and Fermi motion, 
they would nevertheless confirm our expectations for the relic 
neutrino background and give important information about the neutrino 
spectrum. In particular, the $\nu_{e}/\nu_{\mu}$ ratio, shown in 
Figure~\ref{fig:fermimorats}, is a marker for the normal or inverted mass 
hierarchy.


The observation of cosmic-neutrino absorption lines will open the
way---at least in principle---to new insights about neutrino properties
and the thermal history of the universe.  Our calculations, with their
successive inclusion of potentially significant effects, show that how
the tale unfolds will depend on factors we cannot foresee.  The earlier
in redshift the relevant cosmic-neutrino sources appear, the lower the
present-day energy of the absorption lines and the denser the column of
relics the super-high-energy neutrinos must traverse.  In particular,
the appearance of dips at energies much lower that we expect points to
early---presumably nonacceleration---sources, that could give us insight
into fundamental physics at early times and high energy scales.  On the
other hand, integration over a longer range in redshift means more
smearing and distortion of the absorption lines.

If it can be achieved at all, the detection of neutrino absorption
lines will not be done very soon, and the interpretation is likely to
require many waves of observation and analysis.  Nevertheless,
observations of cosmic-neutrino absorption lines offer the possibility
to establish the existence of another relic from the big bang and,
conceivably, they may open a window on periods of the thermal history
of the universe not readily accessible by other means.

\section{Neutrino Coannihilation on Dark-Matter Relics?}
If the nonbaryonic dark matter that makes up some 20\% of the 
mass-energy of the Universe is composed of particle relics, might it 
be possible to observe evidence of neutrino--dark-matter 
coannihilations? Clearly the answer depends on the nature of the 
dark-matter particles. An instructive example is provided by 
neutralino dark matter in the framework of supersymmetric models, so 
I have been analyzing this case with Gabriela Barenboim and Olga 
Mena.\cite{coann}   

If the lightest supersymmetrical particle is the neutralino 
$\chi^{0}_{1}$, with a mass in the neighborhood of $150\gev$, then we 
immediately have two pieces of good news. First, the resonant energy 
for sneutrino formation in the reaction $\nu\chi_{1}^{0} \to 
\tilde{\nu}$ is typically $E_{\nu}^{\mathrm{res}} \approx 400\gev$, 
so we are assured of a reasonable neutrino flux, perhaps even from 
cosmic-ray interactions in the atmosphere.\footnote{The Rome group has 
studied the Lorentz-boosted case of ultrahigh-energy neutralinos 
incident on the relic neutrinos.\cite{Datta:2004sr}} Second, the 
$\snu$-formation cross section is typically about 10\% of the 
$\nu\bar{\nu} \to Z$ annihilation cross section, so is not small.

There ends the good news, at least for the Universe at large.  Whereas
we expect that stable neutrinos should be the most abundant particles
in the Universe after the photons of the cosmic microwave background,
the neutralino gas is on average very tenuous.  The neutralino fraction
of the Universe (identified with the dark-matter fraction) is
$\Omega_{\chi}h^{2} = {\varrho_{\chi}h^{2}}/{\varrho_{c}}$,
where the numerical value of the critical density  is
$\varrho_{c} = 1.05 \times 10^{-5}h^{2}\gev\cm^{-3}$.
Consequently the mass density of relic neutralinos is
$\varrho_{\chi} = 1.05 \times 10^{-5}\cdot  \Omega_{\chi}h^{2} = 
    \bar{\mathcal{N}}_{\chi}M(\chi)$,
where $\bar{\mathcal{N}}_{\chi}$ is the mean number density of relic 
neutralinos throughout the Universe. For neutralino masses in the 150-GeV 
range, $\bar{\mathcal{N}}_{\chi} \ltap
10^{-8}\cm^{-3}$, some ten orders of magnitude smaller than the 
current relic
neutrino density {and 31 orders of magnitude smaller than the 
density of electrons in water.} As a result, the interaction length 
for resonant sneutrino formation is some $10^{15}\mpc$, so 
$\nu\chi\to\snu$ coannihilation is utterly irrelevant in the Universe at 
large.

Our location in the Universe may not be privileged, but it is not
average.  Dark matter clusters in galaxies.  A useful benchmark
is the spherically symmetric universal profile proposed by
Navarro, Frenk, and White,\cite{Navarro:1996he} which yields a
mean dark-matter density (neglecting
local influences) of $\varrho_{\chi} \approx 0.3\gev\cm^{-3}$ at our 
distance from the galactic center, for a number density  $\mathcal{N}_{\chi}
= \hbox{ a
few} \times 10^{-3}\cm^{-3}$, five orders of magnitude enhancement 
over the relic density in the Universe at large. Even considering 
possible enhancements of the relic density in the solar system, the 
rate of interactions produced by atmospheric neutrinos in Earth's 
atmosphere is negligible.

There is one remaining hope, not for neutrino observatories but for 
gamma-ray telescopes. The galaxy as a whole contains some $10^{65}$ 
neutralinos in the scenario we are describing. With a conservative 
neutrino flux, we might expect $10^{24}$ sneutrino events in a year. 
Some fraction of these will decay inelastically and give rise to a 
$\gamma$ spectrum in the few-GeV range. Alas, the number of 
coannihilations viewed by a detector near Earth is only 
$\approx 10^{-21}\cm^{-2}\y^{-1}$.
 
\section{Gravitational Lensing of Neutrinos}
Surely neutrinos---in common with other forms of matter and
energy---experience gravitational interactions.  Where is the
observational evidence to support this assertion?  No analogue of the
classic demonstration of the deflection of starlight by the Sun is in
prospect.  We do not know any continuous intense point source of
neutrinos, and the angular resolution of neutrino
telescopes---{$\approx 5^{\circ}$ for Super-Kamiokande in the
few-MeV range and the $0.5^{\circ}$ goal for km$^{3}$-scale
ultrahigh-energy neutrino telescopes}---is poorly matched to the
anticipated {1.75}-arcsecond deflection of neutrinos from a
distant source.  We must therefore look elsewhere to demonstrate
that neutrinos have normal gravitational interactions.. 

The Supernova 1987A neutrino burst, recorded within 3~h of the
associated optical display after a $166\,000$-year voyage, provides
circumstantial evidence that neutrinos and photons follow the same
trajectories in the gravitational field of our galaxy, to an accuracy
better than 0.5\%.\cite{Longo:1987gc,Krauss:1987me}  On the assumption
that neutrinos do have normal gravitational interactions, weak lensing
induced by relic neutrinos could suppress the large-scale structure
power spectrum on small scales.\cite{Cooray:1999rv,Lesgourgues:2005yv}
The SN1987A argument, though telling, is indirect, and weak lensing by
relic neutrinos has not yet been observed.  Can we imagine a more
direct manifestation of gravity's influence on neutrinos?

Raghavan has recently advocated\cite{Raghavan:2005gn} a neutrino
analogue of the Pound--Rebka experiment\cite{PoundRebka} to demonstrate
the blue shift of neutrinos falling in a gravitational field, applying
the M\"{o}ssbauer effect to recoilless resonant capture of
antineutrinos.  At this meeting, Minakata\cite{Hisa} has suggested that
the method might lead to a tabletop measurement of the neutrino
parameters $\Delta m^{2}_{31}$ and $\theta_{13}$.\cite{Minakata:2006ne}

With my colleagues Olga Mena and Irina Mocioiu, I have been looking
into the possibility of observing the lensing of supernova neutrinos by
the black hole at the galactic center.\cite{nulens} The improbably
ideal case of a supernova explosion on the other side of the galaxy,
symmetric to our position, would be characterized by a prodigious
amplification of the neutrino flux at Earth.  The signals from lensed
supernovae that are not quite so impeccably positioned would be
characterized by noticeable amplification and by dispersion in arrival
time that would stretch the apparent duration of the neutrino burst to
several times its natural length.  We might be richly rewarded for
attending to rare events, for

\begin{quote}
    ``It is a part of probability that many improbable things will
    happen.'' \\ \phantom{MMMMM} --- George Eliot (after Aristotle's 
    \textit{Poetics}), \textit{Daniel Deronda}
\end{quote}

\section{From Neutrino Astronomy to Particle Physics}
As we move into the era of km$^{3}$-scale neutrino
telescopes,\cite{Hulth,Aubert} the baseline strategy of detecting muons
produced by charged-current interactions in or near the instrumented
volume seems certain to achieve the first, astronomical, goal of this 
enterprise: prospecting for distant neutrino sources. In this talk, I 
have tried to show by example how more ambitious detection strategies 
might be rewarded with sensitivity to information of interest to 
particle physics. For the determination of neutrino properties, 
flavor tagging would be immensely valuable, and we can imagine 
obtaining information about neutrino sources and neutrino decays that 
it not otherwise available. The ability to record and to characterize 
the energy of neutral-current events may give decisive sensitivity to 
the onset of new phenomena.

I have also exhibited two reminders of the importance of being
attentive to surprises.  If neutrino sources are associated with
star-bearing regions, and so turn on at redshifts smaller than 20, then
cosmic-neutrino absorption spectroscopy is probably a very distant
dream.  But if non-acceleration sources, such as the decay of cosmic
strings at very early times, are important neutrino producers, then we
might find absorption lines at considerably lower energies than the
rule of thumb $E_{\nu}^{Z\mathrm{res}} = M_{Z}^{2}/2m_{\nu}$ would
suggest.  Gravitational lensing of neutrinos might lead to
characteristic signatures of anomalously intense supernova bursts of
anomalously long duration. Finally, neutrinos interacting with 
dark-matter relics might produce signals for large-area $\gamma$-ray 
telescopes.
  
\section{Acknowledgements}
I salute with pleasure the hard work and perceptive insights of my
extrater\-restrial-neutrino collaborators, Hallsie Reno, Terry Walker,
Raj Gandhi, Ina Sarcevic, Marcela Carena, Magda Lola, Debajyoti
Choudhury, Vu Anh Tuan, Gabriela Barenboim, Olga Mena, and Irina Mocioiu.
Fermilab is operated by Universities Research Association Inc.\ under
Contract No.\ DE-AC02-76CH03000 with the U.S.\ Department of Energy.  I
am grateful for the hospitality of the CERN Particle Theory Group
during the preparation of this talk. Finally, I thank Milla 
Baldo-Ceolin for animating this splendid week in Venice.

\end{document}